\documentclass[prx,twocolumn,english,superscriptaddress]{revtex4-1}
\usepackage[T1]{fontenc}
\usepackage[latin9]{inputenc}
\usepackage{graphicx}
\usepackage{amsmath}  

\makeatletter
\usepackage{babel}
\usepackage{color}

\usepackage{pifont}

\makeatother

\newcommand{\SAVE}[1]{}

\newcommand{\Scal}{{\mathcal S}}

\newcommand{\angstrom}{\mbox{\normalfont\AA}}
\mathchardef\mhyphen="2D 

\begin{document}

%
%
%
	
	\title{Multiphase Magnetism in $\rm Yb_2Ti_2O_7$}
	
	\author{A. Scheie}
	\email{scheieao@ornl.gov}
	\address{Institute for Quantum Matter and Department of Physics and Astronomy, Johns Hopkins University, Baltimore, MD 21218}
	\address{Neutron Scattering Division, Oak Ridge National Laboratory, Oak Ridge, Tennessee 37831, USA}

	\author{J. Kindervater} 
	\address{Institute for Quantum Matter and Department of Physics and Astronomy, Johns Hopkins University, Baltimore, MD 21218}
	
	\author{S. Zhang} %
	\address{Institute for Quantum Matter and Department of Physics and Astronomy, Johns Hopkins University, Baltimore, MD 21218}
	\address{Department of Physics and Astronomy, University of California, Los Angeles, CA 90095}
	
	\author{H.J. Changlani} %
	\address{Department of Physics, Florida State University, Tallahassee, Florida 32306, USA}
	\address{National High Magnetic Field Laboratory, Tallahassee, Florida 32304, USA}
	
	\author{G. Sala} 
	\address{Neutron Spectroscopy Division, Oak Ridge National Laboratory, Oak Ridge, Tennessee 37831, USA}
	
	\author{G. Ehlers} 
	\address{Quantum Condensed Matter Division, Oak Ridge National Laboratory, Oak Ridge, Tennessee 37831, USA}
	
	\author{A. Heinemann} 
	\address{German Engineering Materials Science Centre (GEMS) at Heinz Maier-LeibnitzZentrum (MLZ), Helmholtz-Zentrum Geesthacht GmbH, Lichtenbergstr. 1, D-85748, Garching, Germany}
	
	\author{G. S. Tucker} 
	\address{Laboratory for Neutron Scattering, Paul Scherrer Institut, CH-5232 Villigen, Switzerland}
	\address{Laboratory for Quantum Magnetism, Institute of Physics, Ecole Polytechnique Federale de Lausanne (EPFL), CH-1015 Lausanne, Switzerland}

	\author{S.M. Koohpayeh}  
	\address{Institute for Quantum Matter and Department of Physics and Astronomy, Johns Hopkins University, Baltimore, MD 21218}
	

	\author{C. Broholm}
	\address{Institute for Quantum Matter and Department of Physics and Astronomy, Johns Hopkins University, Baltimore, MD 21218}
	\address{NIST Center for Neutron Research, National Institute of Standards and Technology, Gaithersburg, MD 20899}
	\address{Department of Materials Science and Engineering, Johns Hopkins University, Baltimore, MD 21218}

	\date{\today}

	\begin{abstract}
		We document the coexistence of ferro- and anti-ferromagnetism in pyrochlore $\rm Yb_2Ti_2O_7$ using three neutron scattering techniques on stoichiometric crystals: elastic neutron scattering shows a canted ferromagnetic ground state, neutron scattering shows spin wave excitations from both a ferro-and an antiferro-magnetic state, and  field and temperature dependent small angle neutron scattering reveals the corresponding  anisotropic magnetic domain structure. High-field $\langle 111 \rangle$ spin wave fits show that $\rm Yb_2Ti_2O_7$ is extremely close to an antiferromagnetic phase boundary. Classical Monte Carlo simulations based on the interactions inferrred from high field spin wave measurements confirm $\psi_2$ antiferromagnetism is metastable  within the FM ground state. 
	\end{abstract}
	
	\maketitle

\section{Introduction}

Degeneracy is the foundation for the exotic properties of semimetals \cite{Keimer2017}, fractional quantum hall systems \cite{Keimer2017,Wen1990}, heavy fermion systems \cite{RevModPhys.79.1015},  and  quantum spin liquids \cite{Balents2010review,Savary_2016review,Anderson1973}. Usually, degeneracy is protected by symmetry (such as time reversal symmetry protecting a Kramers doublet), but in mining the vast materials space, we must also expect to encounter non-protected finely tuned degeneracies. Collective phenomena arising from apparently "accidental" degeneracies can share characteristics with symmetry-based counterparts but may also be distinguished by the non-symmetry related nature of the degenerate states. Here we show a remarkably exact but apparently accidental degeneracy exists between ferromagnetism and antiferromagnetism in $\rm Yb_2Ti_2O_7$ and we link it to key collective phenomena of this widely studied quantum magnet.

$\rm Yb_2Ti_2O_7$ is a crystal with magnetic Yb$^{3+}$ ions arranged in a pyrochlore lattice. Its structure and properties were first reported 50 years ago \cite{Blote1969}, but the precise nature of its ground state remains a mystery. It is known to order ferromagnetically at 270 mK \cite{Scheie2017,100HC,FM_order2003,GaudetRoss_order}, but the nature of the ground state is contested \cite{Yaouanc_order,GaudetRoss_order,Bowman2019} and the magnetism is susceptible to disorder at the 1\% level \cite{SeyedPaper,DOrtenzio_noGsOrder,SampleDependence_HC,SampleDependence_Ross,Mostaed2017,Shafieizadeh2018}. What is more, the zero-field neutron spectrum appears to have a diffuse continuum rather than the spin wave excitations expected to accompany ferromagnetism \cite{GaudetRoss_order,Thompson_2017,Ross2009}, and there is evidence of emergent monopole quasiparticles just above the ordering transition \cite{Armitage_monopoles_2016,MonopolesConductivity} whose features may be preserved to lower temperatures in disordered samples \cite{Bowman2019}. The authors' previous work also uncovered a peculiar reentrant field-dependent phase diagram which suggests strong quantum effects \cite{Scheie2017,Changlani2017quantum}.

Many attempts have been made to understand $\rm Yb_2Ti_2O_7$, but a complete account has remained elusive.
Based on fits to high-field spin waves, Ross et al suggested that $\rm Yb_2Ti_2O_7$ may be a quantum spin ice \cite{Ross_Hamiltonian}. However, subsequent studies revising this Hamiltonian \cite{Robert2015,Thompson_2017} showed that $\rm Yb_2Ti_2O_7$ is not a quantum spin ice, but is very close to a phase boundary with the $\psi_3$ antiferromagnetic phase. Theories of multiphase competition between ferromagnetism and antiferromagnetism have been proposed for $\rm Yb_2Ti_2O_7$ \cite{Jaubert2015,Yan2017}, but they have yet to account satisfactorily for the low temperature spectrum, which has been attributed to both multimagnon decay \cite{Thompson_2017} (though this is doubtful given recent theoretical calculations \cite{rau2019magnon}) and fractionalized quasiparticles \cite{chern2018magnetic}.
In short, the nature of the $\rm Yb_2Ti_2O_7$ magnetic ground state remains a matter of debate.

Our progress in understanding $\rm Yb_2Ti_2O_7$ is based on a new class of ultra pure single crystals grown by the traveling solvent floating zone method \cite{SeyedPaper}, the use of the latest high resolution and broad band neutron scattering instrumentation, and modeling using heterogeneous spin wave theory and Monte Carlo Simulation. 
First, we use elastic neutron scattering to refine the ground state magnetic structure and provide evidence for a canted ferromagnetic ground state. Second, we use inelastic neutron scattering to measure the zero-field spin wave spectrum, fit the high-field spectrum to show proximity to the a $\psi_3 + \psi_2$ antiferromagnetic phase, and provide evidence for phase coexistence within the ground state.
Third, we explore magnetic structure on the 100-10000 \AA\ length scale  using small angle neutron scattering, and find a highly anisotropic correlated domain structure.
Fourth, we show using classical Monte Carlo and semiclassical spin wave theory that antiferromagnetism is a metastable state of $\rm Yb_2Ti_2O_7$. 
These findings support the hypothesis that ferromagnetism and antiferromagnetism coexists in $\rm Yb_2Ti_2O_7$  single crystals, forming anisotropic domains that disrupt spin wave propagation. 

\section{Experiments}

\subsection{Elastic Scattering}

\begin{figure}
	\centering\includegraphics[width=0.47\textwidth]{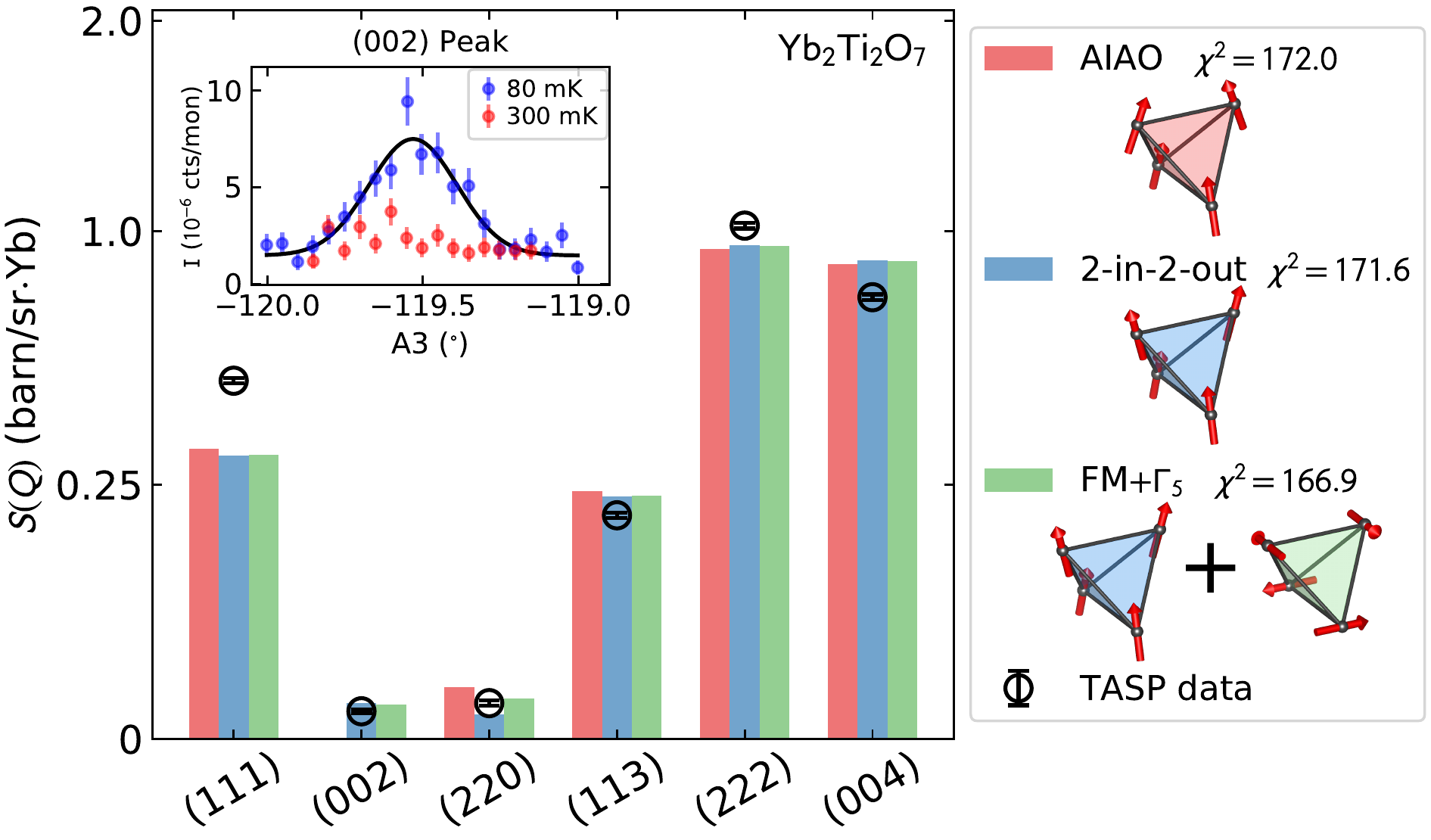}
	
	\caption{Refinement of $\rm Yb_2Ti_2O_7$ magnetic structure based on changes in integrated Bragg intensities upon cooling from 0.6 K to  at 80 mK in zero magnetic field. The inset shows the zero-field intensity of the $(002)$ Bragg peak appears upon cooling from 300 mK to 80 mK, which rules out AIAO order. 
		Note that the $S(Q)$ axis is quadratic.}
	\label{flo:Refinement}
\end{figure}

We collected $\rm Yb_2Ti_2O_7$ single-crystal elastic scattering data on the TASP triple axis spectrometer at PSI. The sample was a 0.4 g sphere cut from a high quality $\rm Yb_2Ti_2O_7$ single crystal (the same sample as in ref. \cite{Scheie2017}). The sphere was mounted on a copper finger, and loaded in a dilution refrigerator insert with the $\langle 110 \rangle$ direction vertical. We collected elastic scattering ($\hbar \omega = 0.0 \pm 0.1$ meV) data on the $(1 \bar 1 1)$, $(002)$, $(2 \bar 2 0)$, $(1 \bar 1 3)$, $(2 \bar 2 2 )$, and $(004)$ Bragg peaks, tracking the peak area as a function of temperature. We discovered significant multiple scattering (particularly on the $(002)$ peak), so we collected data using both 5 meV and 4.5 meV neutrons to isolate single event diffraction (see appendix \ref{app:ElasticScattering} for details). 
We converted the peak intensity to absolute units using the fraction of the accompanying nuclear peak intensity (using nuclear structure factors calculated from the structure reported in ref. \cite{SeyedPaper}. For $(002)$ we interpolated the normalization constant inferred from other Bragg peaks as $(002)$ has zero nuclear intensity), and fit to theoretical spin structure models assuming equal population of domains. The results are shown in Fig. \ref{flo:Refinement}.

\subsection{Spin wave spectrum}

We mapped the field and temperature dependent energy-resolved neutron scattering cross section of $\rm Yb_2Ti_2O_7$ on the CNCS spectrometer at ORNL.
We co-aligned three stoichiometric $\rm Yb_2Ti_2O_7$ crystals totaling 7.2 g oriented with the $\langle 111 \rangle$ direction along a vertical magnetic field and mounted in a dilution refrigerator. (These crystals were grown by the same technique as the sphere \cite{SeyedPaper}.)
We collected neutron scattering data with two instrument configurations. We did a high-neutron-flux measurement with $E_i=3.32\>{\rm meV}$ neutrons, the Fermi chopper at 60 Hz, and the disk chopper at 300 Hz for an energy resolution of $0.15$ meV full width at half maximum (FWHM). We also performed a higher resolution measurement with $E_i=3.32\>{\rm meV}$ neutrons, a Fermi chopper at 180 Hz, and disk chopper at 240 Hz for a FWHM energy resolution of $0.08$ meV.

We collected data at 0 T (both from a field-cooled and a zero-field-cooled state), 0.35 T, 0.7 T, and 1.5 T at base temperature (100 mK) in the high flux configuration, and then zero-field-cooled 0 T data in the high resolution configuration. We also collected scattering data at 20 K for use as a background. We subtracted this background from all data sets, divided by the squared Yb$^{3+}$ magnetic form factor, symmetrized the data using  Mantid \cite{Mantid}, and made cuts along high-symmetry directions $\Gamma \> (000) \rightarrow \rm K \> (2 \bar 2 0) \rightarrow UL \> (2 \bar 1 1) \rightarrow \Gamma \> (000)$.
All high-symmetry cuts plotted in this paper (such as Fig. \ref{flo:SW_comparison}) are symmetrized, but all plotted constant-energy slices [such as Fig. \ref{flo:SW_FM-AFM}(e)] are unsymmetrized.

To compare the data to theoretical models, we used the SpinW package \cite{SpinW} to simulate the inelastic scattering cross section based on linear spin wave theory and the Hamiltonians in refs. \cite{Ross_Hamiltonian,Robert2015,Thompson_2017}. We fitted the $1.5 \> {\rm T}$ data (where the spin wave modes are most clearly defined) to a linear spin wave model, and extracted a new set of parameters.
The data compared with previous theoretical studies is plotted in Fig. \ref{flo:SW_comparison}, and the best fit spin wave model is plotted in Fig. \ref{flo:BestFit}. While consistent with previous scattering data, the revised exchange constants are needed to account for our new data acquired in a different reciprocal lattice plane.

\subsection{Small angle neutron scattering}

We performed two small angle neutron scattering (SANS) experiments on the SANS-1 instrument at MLZ FRM-II. Both experiments used the same 0.4 g spherical sample of $\rm Yb_2Ti_2O_7$  in a dilution refrigerator, and oriented with the $\langle 110 \rangle$ direction vertical and the $\langle 111 \rangle$ direction along a horizontal magnetic field oriented perpendicular to the neutron beam. We collected data both with 4.6 \AA \space and 17 \AA \space neutrons, for access to wave vector transfer from 0.001 \AA$^{-1}$ to 0.023 \AA$^{-1}$.

We acquired SANS data for temperatures between 800 mK and 70 mK and fields from 0 T to 1 T. The basic scattering pattern is shown in Fig. \ref{flo:SANS_FC-ZFC}, the field and temperature dependence of SANS features are plotted in Fig. \ref{flo:SANS_B-T}, and the $Q$ dependence is presented in Fig. \ref{flo:SANS_Porod}.

\section{Elastic Scattering} \label{sec:Diffraction}

The measured magnetic Bragg peak intensities are compared to three models in Fig. \ref{flo:Refinement}: a ferromagnetic all-in-all-out (AIAO) structure (proposed in refs. \cite{Yaouanc_order,Bowman2019}), a 2-in-2-out canted ferromagnetic structure (proposed in refs. \cite{GaudetRoss_order,Yan2017}), and a combination of a ferromagnetic phase and an antiferromagnetic phase FM+$\Gamma_5$. (The two states $\psi_3$ and $\psi_2$ within $\Gamma_5$ have equivalent Bragg intensities for unpolarized neutrons and averaging over all domains, so we are not able to distinguish between them here.) 
No refinement matches the data perfectly, but the best fit is the FM+$\Gamma_5$, with an ordered moment of 1.3(1) $\mu_B$, a canting angle of 5.3(8)$^{\circ}$, and an effective antiferromagnetic moment of 0.1(2) $\mu_B$. 
The large uncertainty on the antiferromagnetic phase is consistent with there being no long-range $\psi_3$ order. 
The $\psi_3$ phase has large $(220)$ Bragg intensity but the observed $(220)$ peak is quite small, which constrains the $\Gamma_5$ AFM phase to be $< 10$\%  of the crystal based on Bragg intensities (diffuse scattering from short-range or quasi-two-dimensional order is another story as it will not contribute significantly to coherent Bragg diffraction). 

The presence of an $(002)$ magnetic Bragg peak has been a matter of some controversy between different diffraction studies \cite{Yaouanc_order,Bowman2019,Antonio2017}, and it is the key difference between the proposed AIAO splayed ferromagnetic and 2-in-2-out splayed ferromagnetic structures. As shown in the inset to Fig. \ref{flo:Refinement}, we definitively detect a nonzero magnetic intensity at $(002)$ and we can exclude multiple scattering as the origin (see Appendix \ref{app:ElasticScattering}). This  rules out the AIAO structure in the present sample. 
This is consistent with  the famous pyrochlore phase diagram in ref. \cite{Yan2017} (which does not contain the proposed AIAO order) and the associated model Hamiltonian. Possible reasons for the less than perfect fit (in particular magnetic diffraction at (111) is weaker than anticipated) is unequal domain population (the mild clamping force applied to the sample does break cubic symetry) and extinction\cite{SteffenPaper}. 


\section{Spin Wave Spectrum}  \label{sec:SpinWaves}

\begin{figure}
	\centering\includegraphics[scale=0.49]{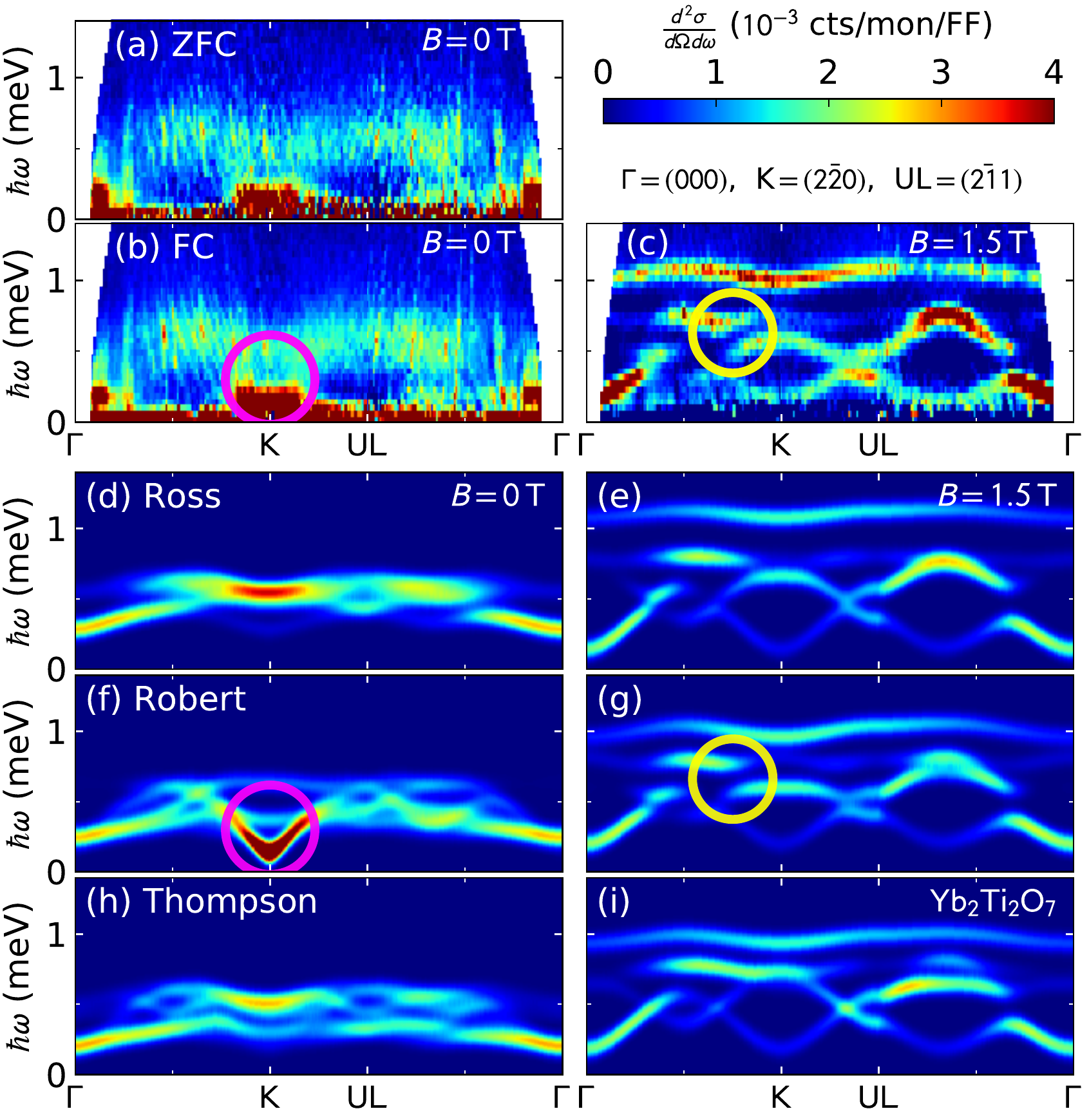}
	
	\caption{Inelastic magnetic neutron scattering from $\rm Yb_2Ti_2O_7$ in the (a) $B=0$ zero field-cooled (ZFC) state, (b) $B=0$ field-cooled (FC) state, (c) $B=1.5 \> {\rm T}$, compared to linear spin wave theory predictions from (d)-(e) Ross et al's Hamiltonian \cite{Ross_Hamiltonian}, (f)-(g) Robert et al's Hamiltonian \cite{Robert2015}, and (h)-(i) Thompson et al's Hamiltonian \cite{Thompson_2017}.  The vertical streaks in the zero-field data near $\Gamma$ are due to saturated neutron detectors.
	}
	\label{flo:SW_comparison}
\end{figure}
\begin{figure*}
	\centering\includegraphics[scale=0.42]{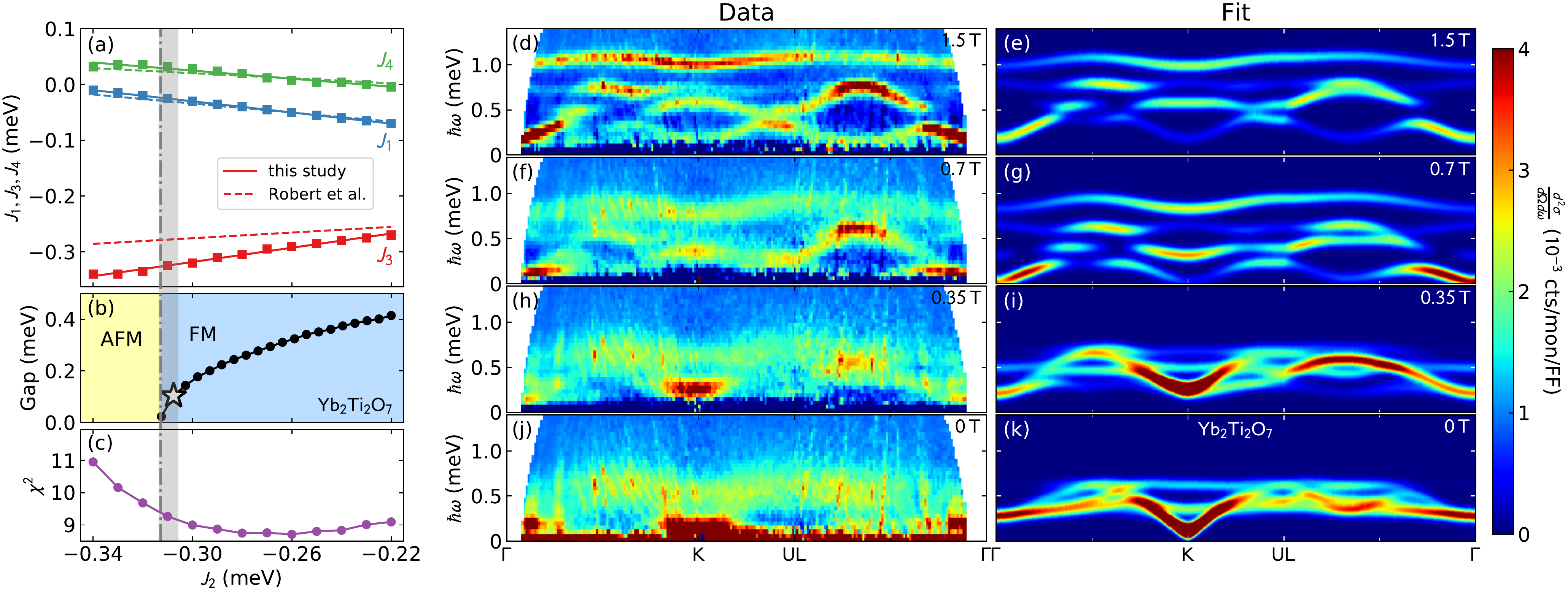}[H]
	
	\caption{Fit to $\rm Yb_2Ti_2O_7$ spin wave spectrum. (a) Line through parameter space minimizing $\chi^2$ of the calculated vs observed intensity data at 1.5 T. (b) Spin wave gap at $(220)$ plotted against $J_2$, with the star indicating a gap of 0.11 meV. The gap goes to zero at the phase boundary between the FM and AFM $\psi_3$ phase. (c) $\chi^2$ along the "best fit line" in panel (a). 
		Panels (d)-(k) show the measured neutron scattering data and the calculated neutron scattering cross section for the Hamiltonian fitted to the 1.5 T data. The experimental plots are averaged over $\pm 0.038$ \AA$^{-1}$ in the direction perpendicular to the cut.
	}
	\label{flo:BestFit}
\end{figure*}

The inelastic magnetic neutron scattering cross section  (plotted in Figs. \ref{flo:SW_comparison} and \ref{flo:BestFit}) contains broadened spin-wave-like ridges in zero-field, and sharper spin-wave-like modes in a field of 1.5 T applied along $\langle 111 \rangle$.
This result is significant because previous studies in crystals grown at higher temperatures without the benefit of a traveling solvent  did not detect spin-wave-like scattering in zero field \cite{Ross_Hamiltonian,Thompson_2017,GaudetRoss_order,Ross2009}; presumably the higher quality crystal in this experiment makes the difference.
There is no visual difference between the field-cooled and zero-field cooled spectrum  that we detect [Fig. \ref{flo:SW_comparison}(a)-(b)] except a slight enhancement of intensity in the FC case. In zero-field, the modes appear to come down to the elastic line at the $(220)$ Bragg peak.

Having observed zero field spin waves, albeit broadened, we can compare the observations to the magnetic scattering cross section associated with the Hamiltonians reported by Ross et al \cite{Ross_Hamiltonian}, Robert et al \cite{Robert2015}, Thompson et al \cite{Thompson_2017} (hereafter referred to as the "Ross", "Robert", and "Thompson" Hamiltonians, with the same $g$ tensors as used in each respective study) based on linear spin wave theory. We used the SpinW package \cite{SpinW} to carry out these calculations. 
As shown in Fig. \ref{flo:SW_comparison}(d), (f), and (h), the only Hamiltonian which predicts a soft (nearly gapless) mode at $(220)$ is Robert; the rest predict a gapped spectrum.
At 1.5 T, we again find that the Robert Hamiltonian yields the best match to the data. In particular, the shape of the modes at $(1.5,1.5,0)$ do not match calculations based on the Thompson and Ross Hamiltonians well, but they do match the Robert predictions well.
The agreement with the Robert Hamiltonian spin waves is not perfect---for example, the highest $\Gamma$ point mode at 1.5 T is at 0.9 meV instead of 1.0 meV as observed---but these comparisons make it clear that the Robert Hamiltonian treated with lowest order spin wave theory is the closest to describing the zero field broad spin wave modes in  $\rm Yb_2Ti_2O_7$. (As an aside, this comparison highlights the need to be careful with high-field fits to spin wave modes as they can be underconstrained.)

To further refine the magnetic Hamiltonians, we fit
the nearest neighbor exchange matrix to the corresponding intensity of the 1.5 T neutron spectrum where the spin wave modes are sharpest. For Yb on the pyrochlore lattice the exchange matrix takes the form
\begin{equation}
J = \left( \begin{matrix} 
J_2 & J_4 & J_4 \\ -J_4 & J_1 & J_3 \\ -J_4 & J_3 & J_1
\end{matrix} \right),
\end{equation}
where $J_1$ (XY), $J_2$ (Ising), $J_3$ (pseudo-dipolar), and $J_4$ (DM) are the four symmetry-allowed exchange variables \cite{Ross_Hamiltonian}. $\chi^2$ is defined by subtracting intensity pixel-by-pixel from a simulated data set, thereby fitting both the  energy and intensity of the modes. (We used the $g$ tensor from ref. \cite{Thompson_2017}; see appendix \ref{app:SpinWaveFits} for details). Just like Robert et al's refinement of Ross et al's data \cite{Robert2015}, we find that there is a "best fit line" through parameter space ($J_1$, $J_2$, $J_3$, $J_4$) where the $\chi^2$ value is approximately constant. This is depicted in Fig. \ref{flo:BestFit}(a), with $\chi^2$ in panel (c). Our line is nearly the same as in Robert et al \cite{Robert2015}.
The fit can be constrained to a point along this line by considering the $(220)$ spin wave gap.

The zero-field gap at $(220)$ is a measure of proximity to the $\psi_3$ AFM phase. Spin wave simulations show that as the Hamiltonian approaches the AFM ordered phase, soft spin wave modes come down in energy close to the $(220)$, $(111)$, and $(113)$ Bragg peaks (which are the magnetic Bragg peaks for $\psi_3$), until they touch zero energy right at the phase boundary (see appendix \ref{app:SoftModes} for details). So the smaller the gap, the closer to the FM+AFM phase boundary.
High resolution spin wave measurements in Fig. \ref{flo:SW_gap}(a) show a gap around  0.11(3) meV (see Discussion section), which is slightly smaller than 0.136 meV, obtained from Robert's Hamiltonian.
We can use this knowledge of the gap size to constrain the fit along the best fit line, as shown in Fig. \ref{flo:BestFit}(b). This yields the following optimized parameters
\begin{equation}
\begin{split}
J_1 = -0.026(2) \> {\rm meV} \quad & \quad
J_2 = -0.307(3) \> {\rm meV} \\
J_3 = -0.322(3) \> {\rm meV} \quad & \quad
J_4 = 0.028(2) \> {\rm meV}
\end{split}.
\label{eq:FittedH}
\end{equation}
For a comparison to previously published results see table \ref{tab:parameters}. 

\begin{figure}
	\centering\includegraphics[scale=0.42]{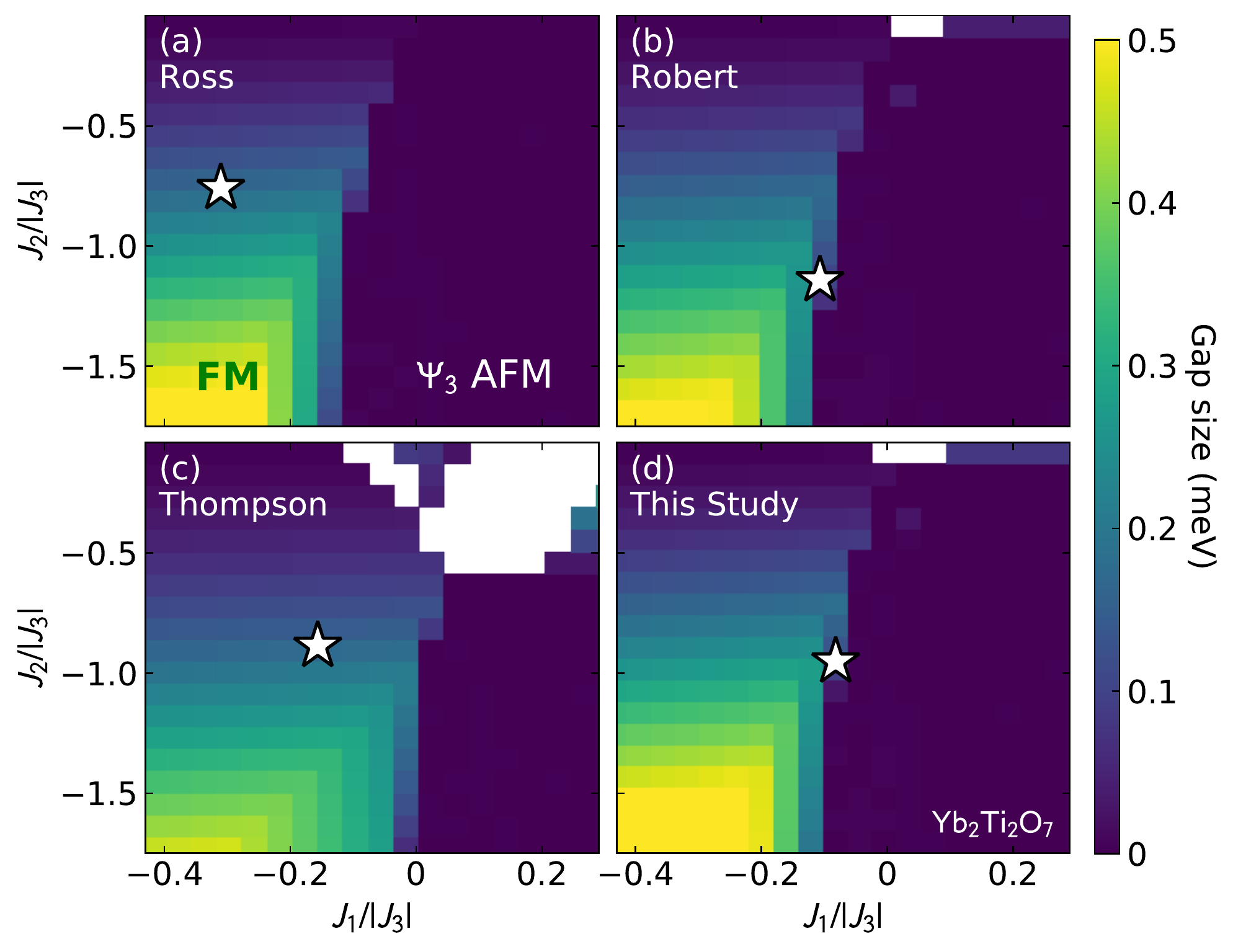}
	
	\caption{Calculated zero-field spin wave gap at $(220)$ for $\rm Yb_2Ti_2O_7$ as a function of $J_1$ and $J_2$ for four Hamiltonians: (a) Ross et al \cite{Ross_Hamiltonian}, (b) Robert et al \cite{Robert2015}, (c) Thompson et al \cite{Thompson_2017}, and (d) this study. ($J_3$ and $J_4$ are held fixed. White areas are where neither the FM nor $\psi_3$ AFM are stabilized.) The gap goes to zero upon entering the $\psi_3$ AFM ordered phase, and this study places $\rm Yb_2Ti_2O_7$ closer to the AFM phase boundary than the other three Hamiltonians.
	}
	\label{flo:PhaseDiagram}
\end{figure}

To show how close this Hamiltonian is to the phase boundary, we built a phase diagram by mapping the calculated $(220)$ gap as a function of $J_1/|J_3|$ and $J_2/|J_3|$  for the four different Hamiltonians (holding $g$ and $J_4$ constant), shown in Fig. \ref{flo:PhaseDiagram}. This study puts $\rm Yb_2Ti_2O_7$ closer to the FM+AFM phase boundary than any previous experimental study, though the Robert Hamiltonian also comes close. In passing, we note that $J_4$, although neglected in the plotted phase diagram of ref. \cite{Yan2017}, noticeably shifts the phase boundary between the FM and AFM states.


\subsection*{AFM+FM phase coexistence}

Although linear spin wave theory based on the fitted Hamiltonian successfully reproduces many features in the $\rm Yb_2Ti_2O_7$ neutron spectrum, there are some features which it does not reproduce, most notably the broadened excitations at low magnetic fields. We shall argue that these features arise from FM+AFM phase coexistence. 

\begin{figure}
	\centering\includegraphics[scale=0.49]{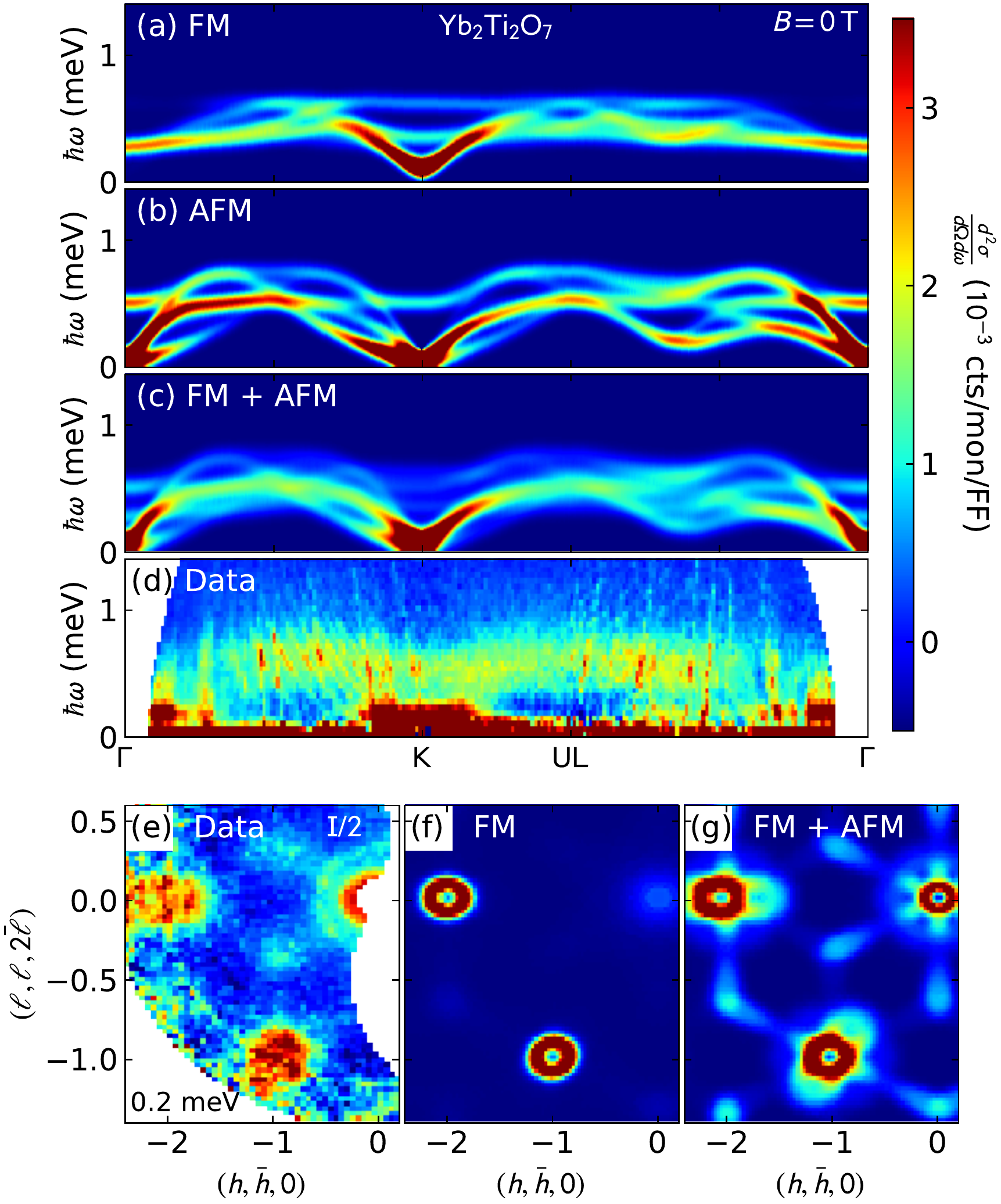}
	
	\caption{Inelastic magnetic neutron scattering data for $\rm Yb_2Ti_2O_7$ compared to the calculated magnetic neutron scattering cross section for ferromagnetic (FM) and antiferromagnetic (AFM) and mixed FM+AFM ground states. (a) Calculated magnetic neutron scattering cross section for the best fit Hamiltonian ferromagnetic (FM) ground state. (b) Calculated spectrum for an antiferromagnetic $\psi_3$ (AFM) state just across the phase boundary. (c) Average of FM and AFM spectra. (d) Field-cooled 70 mK spectrum. (e) Constant 0.2 meV energy slice for $\rm Yb_2Ti_2O_7$ (FC) at 70 mK scattering. (f) and (g) show the 0.2 meV spin wave theory calculations for the FM and FM+AFM states, respectively. The overall scattering matches the AFM + FM result better, particularly near $(- \frac{4}{3}, \frac{2}{3}, - \frac{2}{3})$ at 0.2 meV.
	}
	\label{flo:SW_FM-AFM}
\end{figure}

Within the broadened zero field excitations (which may be broadened by multimagnon decay \cite{Thompson_2017} or by a heterogeneous ground state), a feature not captured by our fitted Hamiltonian is that the zero-field spectrum at $(- \frac{4}{3}, \frac{2}{3}, - \frac{2}{3})$ (between UL and $\Gamma$) appears to come close to zero-energy [see Fig. \ref{flo:BestFit}(j)-(k)]. The same is true of the low-energy scattering near $\Gamma$.
The ferromagnetic spectrum does not reproduce these features. However, the Hamiltonian can be "nudged" into the AFM phase (by setting $J_2 = -0.313$ meV), and then the simulated scattering spectrum does reproduce these features (see Fig. \ref{flo:SW_FM-AFM}). If one takes the average of the FM and AFM spectra (50\% weight on each), the simulated spectrum resembles the experimental data more closely (Fig. \ref{flo:SW_FM-AFM}(c)). Fitting the ratio between FM and AFM scattering with the high-symmetry cuts gives 43(3)\% AFM and 57(3)\% FM (see Fig. \ref{SI:FM-AFM-fit}).
The improvement over the FM spectrum becomes especially evident when comparing constant energy slices, as shown in Fig. \ref{flo:SW_FM-AFM}(e)-(g). The lobe of scattering at $(- \frac{4}{3}, \frac{2}{3}, - \frac{2}{3})$ observed in the experimental data is absent in the FM calculated spectrum, but it is clearly present in the FM+AFM spectrum. In addition, the FM+AFM spectrum reproduces the elongated spin wave dispersions above the $(220)$ peaks. 
These features indicate the presence of the $\psi_3$ AFM phase in the zero-field ground state of $\rm Yb_2Ti_2O_7$.

If this line of reasoning is correct, the observed broadening is partly due to the overlap of many spin wave modes [see Fig. \ref{flo:SW_FM-AFM}(c)]. However, it only seems to apply to the zero-field state: at nonzero magnetic fields the FM+AFM spectrum does not match the scattering data (see Fig. \ref{flo:SI_FM-AFM} in Appendix \ref{app:SpinWaveFits}). Thus it appears that the FM+AFM phases only coexist at very low-fields ($|B|<0.1 \> {\rm T}$).

One may object that the above analysis is inconsistent with the elastic scattering, which shows the AFM state as $<10$\% of the ground state order. However, it should be noted that the Bragg peaks are exclusively associated with three dimensional long range order. If the AFM order were present in smaller regions (for instance, within domain walls) or on short timescales, it would not produce sharp Bragg peaks and would only weakly influence the refinement of the Bragg diffraction data. It is possible for confined AFM to be present without sharp AFM Bragg peaks.

\section{Theoretical Simulations}

To better understand the nature of the coexisting AFM and FM phases, we simulated the neutron spectrum of $\rm Yb_2Ti_2O_7$ by first performing classical Monte Carlo (MC) simulation based on our spin Hamiltonian in Eq. \ref{eq:FittedH}, followed  by 
linear spin wave theory on the corresponding optimized spin configuration. The MC simulations were conducted for $T=0.2$~K using single spin updates for continuous spin on pyrochlore lattices (with 16 site cubic unit cells) of size $N=16 L^3$ for $L=8$ (8192 total spins) for a total of $2 \times 10^8$ steps. Further iterative minimization~\cite{lapa_henley} was performed on the last configuration encountered in the run to obtain the classical spin configuration that corresponds to the nearest local energy minimum. 
This entire process was performed 400 times for different starting random seeds. Such a simulation cell is found to not be large enough to capture domain effects, but it is large enough to probe the stability of AFM and FM states.
We found that each of these configuration became trapped either in the cubic ferromagnetic phase or in the $\psi_2$ + $\psi_3$ manifold with a preference for $\psi_2$ (Fig. \ref{flo:MC_results}). In all 400 cases, the configurations corresponds to a $q=0$ state i.e. all tetrahedra were alike. This lends credence to the idea of phase coexistence: classically, pockets of the system are kinetically trapped in the $\psi_2$ AFM phase. 

The preference for $\psi_2$ is surprising because the $\rm Yb_2Ti_2O_7$ Hamiltonian is near the $\psi_3$ phase boundary. However, because the lowest energy $\rm Yb_2Ti_2O_7$ spin configuration is a canted ferromagnet, the $\psi_3$ is an unstable saddle-point in energy and will rotate into a ferromagnetic ground state without increasing the total energy of the spin configuration \cite{Yan2017}. $\psi_2$, meanwhile, is protected by a finite energy barrier and becomes favored by fluctuations (see Appendix \ref{app:MonteCarloGroundStates} for details). Thus, while $\psi_2+\psi_3$ is an exactly degenerate manifold, we do not expect to see any stable $\psi_3$ states, which is consistent with our MC results.

\begin{figure}
	\centering\includegraphics[scale=0.49]{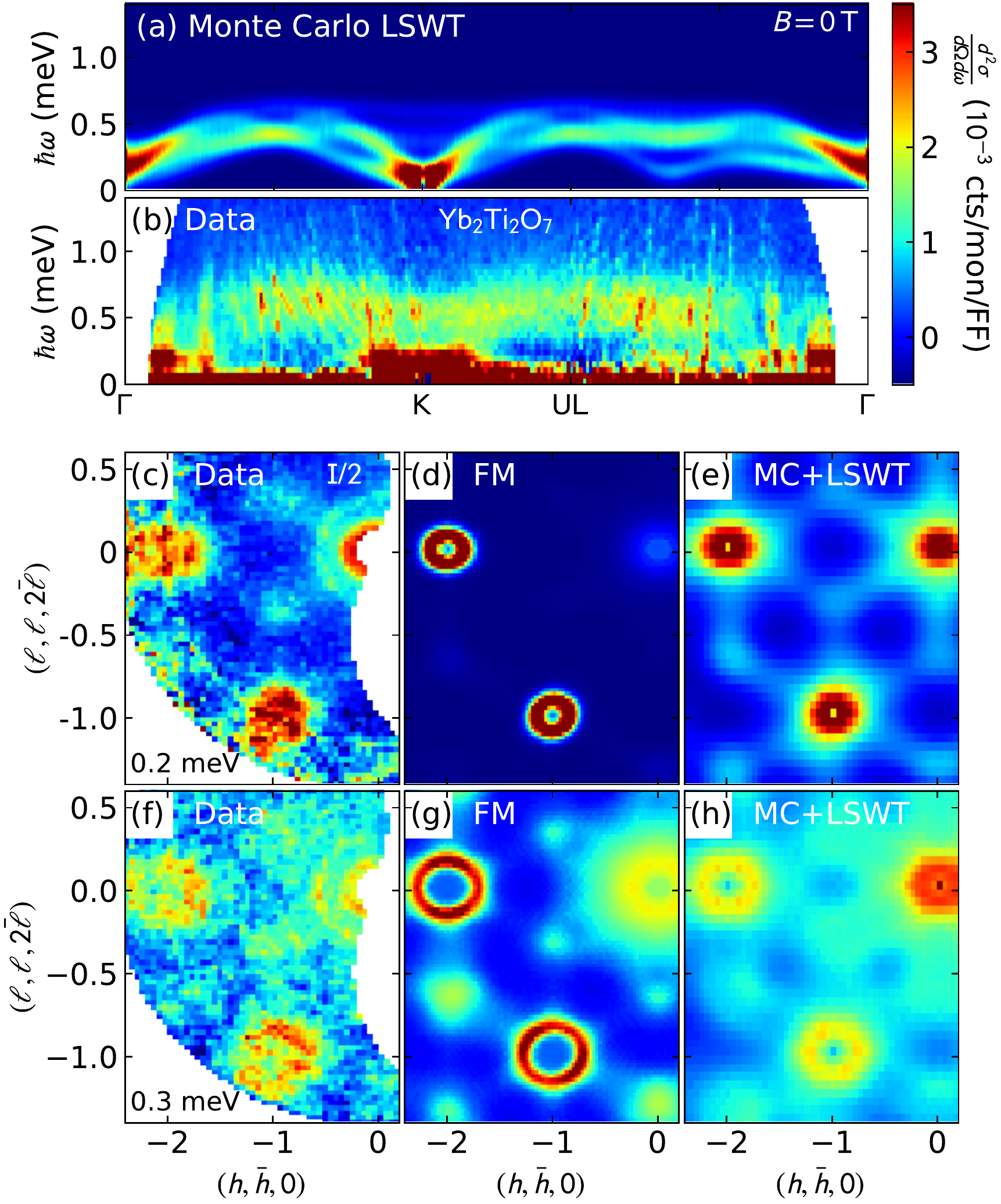}
	
	\caption{(a) MC+LSWT calculated neutron spectrum for $\rm Yb_2Ti_2O_7$ compared to the (b) Field-cooled 70 mK spectrum. (c)-(h) show constant energy (0.2 and 0.3 meV) slices for $\rm Yb_2Ti_2O_7$ (FC) at 70 mK scattering compared to the FM and MC+LSWT spectra. The MC+LSWT simulation better matches the scattering pattern, particularly near $(- \frac{4}{3}, \frac{2}{3}, - \frac{2}{3})$ at 0.2 meV.
	}
	\label{flo:SW_FM-AFM-MC}
\end{figure}

Calculating and summing the calculated spin-wave spectra from all 400 states (computational details are in Appendix \ref{app:SemiclassicalMC}), we get a remarkable (though imperfect) agreement with the observed spin wave spectrum, as shown in Fig. \ref{flo:SW_FM-AFM-MC}. 
This suggests the broadened zero-field spin wave spectrum arises from admixture of antiferromagnetic regimes into the otherwise ferromagnetic low-$T$ spin configuration.

\begin{figure}
	\centering\includegraphics[scale=0.43]{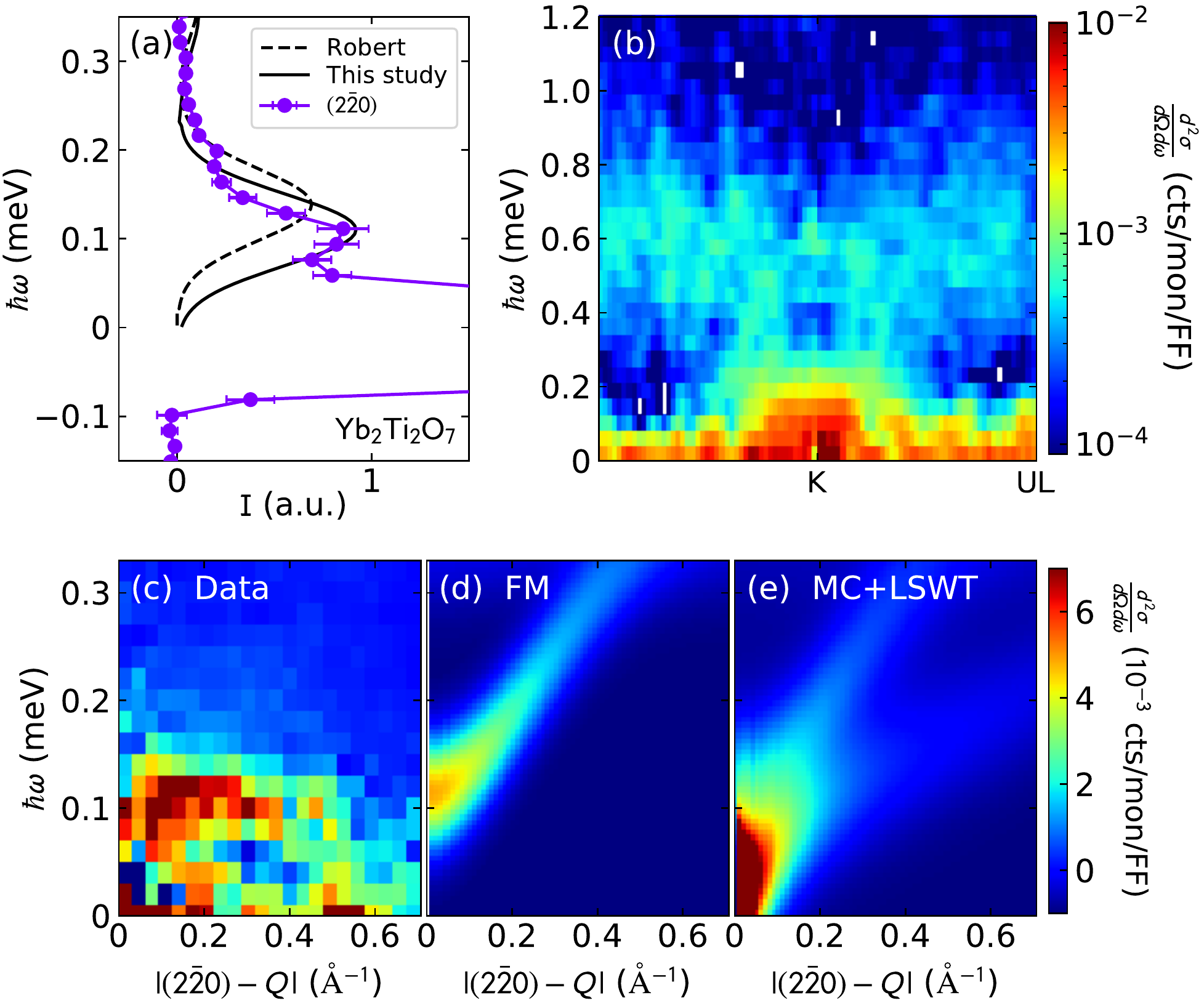}
	
	\caption{(a) Spin wave gap at $(2 \bar20)$ for $\rm Yb_2Ti_2O_7$ at 100 mK, plotted with the predicted gap from Robert  \cite{Robert2015} and this study. (b) Zero-field spin wave spectrum around $(2 \bar20)$ (following the same cuts as Fig. \ref{flo:BestFit}) measured at higher resolution. (c) Spin wave spectrum plotted against distance from $(2 \bar20)$. (d)-(e) Calculated spin wave spectrum plotted against distance from $(2 \bar20)$ for (d) a ferromagnetic ground state and (e) the MC+LSWT spectrum. Neither reproduce the low-lying flat feature above $(2 \bar20)$.
	}
	\label{flo:SW_gap}
\end{figure}

Although the phase coexistence hypothesis successfully accounts for many features in the measured neutron spectrum, not all features are captured by the MC+LSWT linear spin wave simulations. Most notable is the low-lying excitation above the $(220)$ Bragg peak shown in Fig. \ref{flo:SW_gap} [circular binning around $(220)$  is shown in Fig. \ref{flo:SW_gap}(c)-(e) to highlight the dispersion].  Spin wave theory, both from the FM and AFM phase, predict strongly dispersive modes atop $(220)$. However, the measured spectrum shows an intense flat mode at 0.11 meV that is not accounted for by any of these simulations.
A similar flat mode was also observed above the $(111)$ Bragg peak by Antonio et al \cite{Antonio2017} at 0.10 meV. These may possibly be associated with tunneling in and out of the AFM phase.  Alternatively standing AFM spin waves within a finite size AFM domain or domain wall might produce such scattering.
Because of the apparent correspondence with the soft spin wave modes, we anticipate that such a low-energy flat mode will also be found above the $(113)$ Bragg peak.

%


Considering regions of AFM within a FM ground state brings us to larger length scales where dipolar interactions become relevant.
To experimentally probe such length scales and explore the role of magnetic domains and dipolar interactions in $\rm Yb_2Ti_2O_7$, we turn to small angle neutron scattering.

\section{Small angle neutron scattering} \label{sec:SANS}

\begin{figure}
	\centering\includegraphics[scale=0.45]{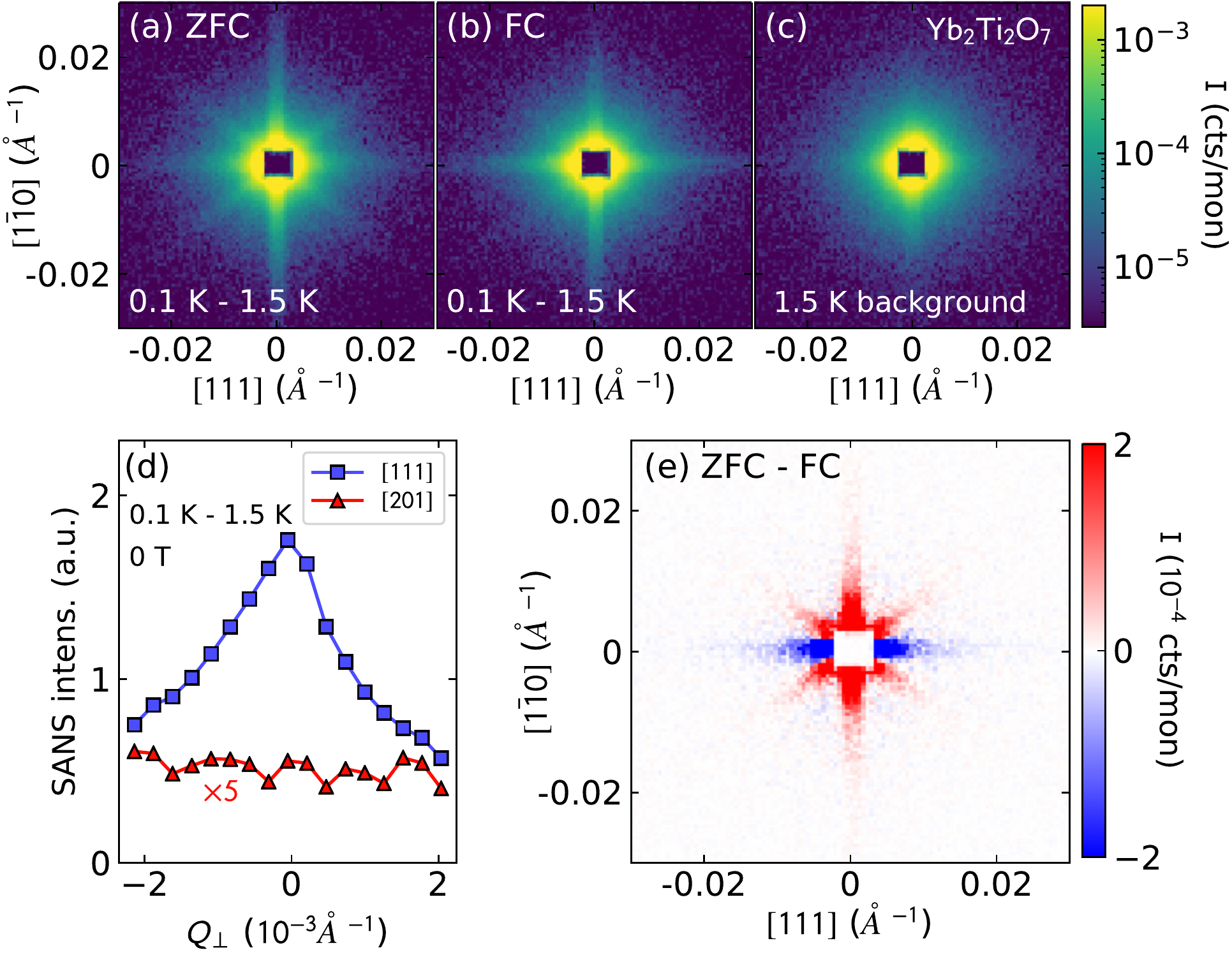}
	
	\caption{SANS data for $\rm Yb_2Ti_2O_7$ in (a) the zero-field-cooled state, and (b) the field-cooled state after applying a 1 T field in the $\langle 111 \rangle$ direction and subtracting (c) a 1.5 K background. (d) Dependence of the $(111)$ and $(201)$ scattering streaks on wave vector transfer $Q_{\perp}$ perpendicular to the plane of the detector.  with 1.5 K data subtracted. Ranges of $Q$ integrated are $Q_{111} = 0.015\pm0.011$~\AA$^{-1}$ and $Q_{201} = 0.016\pm0.011$~\AA$^{-1}$ shown in Fig. \ref{flo:SANS_Porod}. (e) Difference between panels (a) and (b), revealing an enhancement of $(111)$ scattering rod.
	}
	\label{flo:SANS_FC-ZFC}
\end{figure}

\begin{figure}
	\centering\includegraphics[scale=0.38]{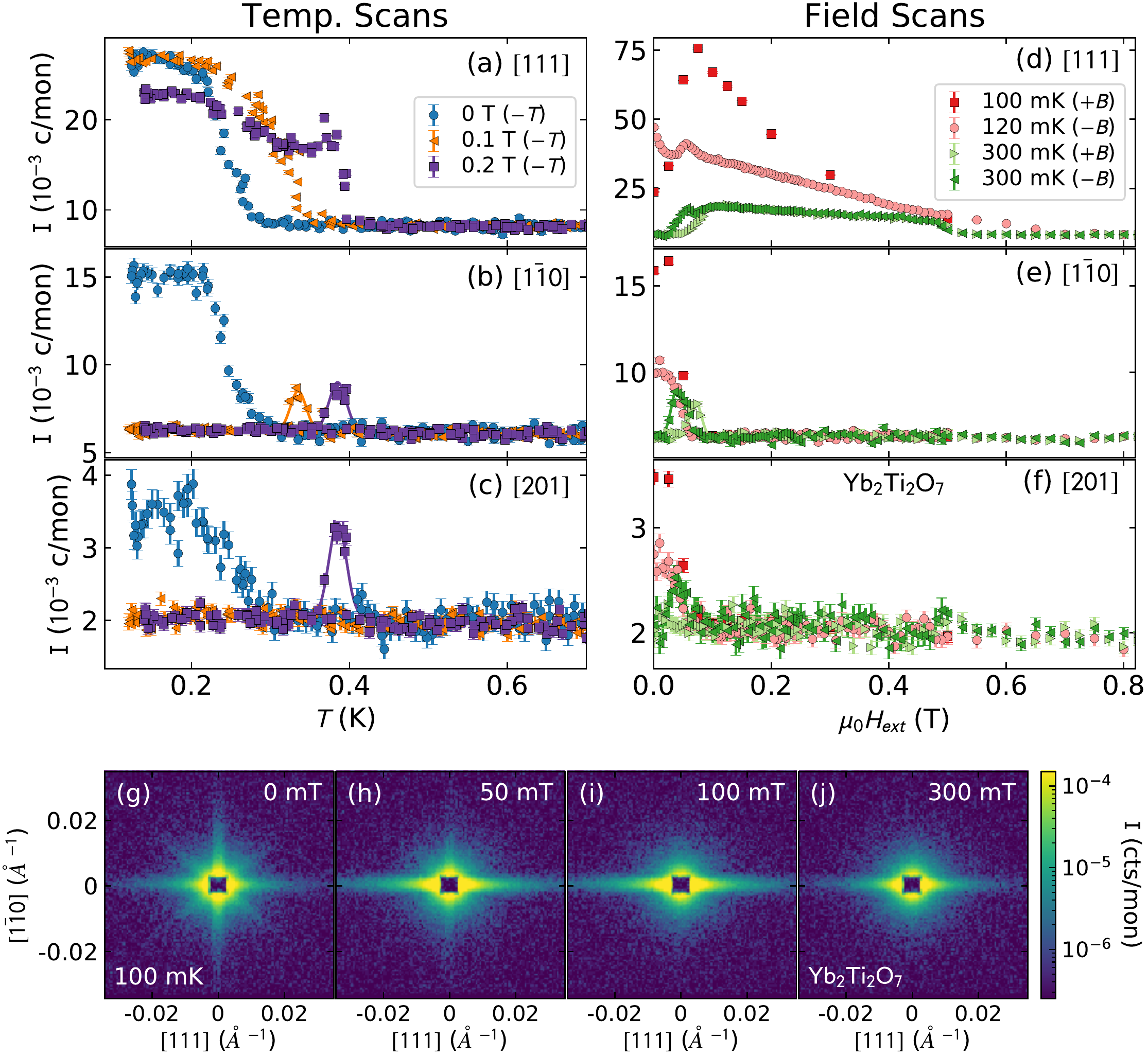}
	
	\caption{Field and temperature dependent SANS data for $\rm Yb_2Ti_2O_7$. (a)-(c) show temperature dependence of the (a) $(111)$, (a) $(1 \bar 1 0)$, and (a) $(201)$ rods at 0T, 0.1 T, and 0.2 T. (d)-(f) show field dependence of the (d) $(111)$, (e) $(1 \bar 1 0)$, and (f) $(201)$ rods at 120 mK and 300 mK. The $(111)$, $(1 \bar 1 0)$, and $(201)$ directions are defined in the insets to Fig. \ref{flo:SANS_Porod}(a)-(c). 
		(g)-(j) show colormap images of the $\rm Yb_2Ti_2O_7$ SANS pattern as a function of field at 100 mK.
	}
	\label{flo:SANS_B-T}
\end{figure}

Within the magnetic ordered phase, the small angle neutron scattering (SANS) pattern shown in Fig. \ref{flo:SANS_FC-ZFC} has a star pattern formed by streaks of scattering extending along the $(111)$, $(1 \bar10)$, and $(201)$ directions.  The $(111)$ streaks  have been previously reported \cite{Buhariwalla2018}, but not the streaks of scattering along the $(1 \bar10)$ and $(201)$ directions.
The SANS disappears above the magnetic ordering temperature of 270 mK [Fig. \ref{flo:SANS_FC-ZFC}(c) and Fig. \ref{flo:SANS_B-T}(d)-(f)], which suggests it is associated with magnetic order. The long length scale ($\frac{2\pi}{0.006 \>\angstrom^{-1}} \approx 1000$ \angstrom), suggests this scattering is  associated with a magnetic domain structure within $\rm Yb_2Ti_2O_7$. The SANS data are consistent with faceted domain walls with particular orientations along high-symmetry directions and no long-range pattern.

The SANS scattering is different in the field-cooled (FC) and zero-field-cooled (ZFC) states. Fig. \ref{flo:SANS_FC-ZFC}(e) shows the scattering along  $( 111 )$ is greatly enhanced after applying a  $\langle 111 \rangle$ magnetic field. This indicates that domain walls that extend perpendicular to $\langle 111 \rangle$ proliferate after reducing a $\langle 111 \rangle$ oriented field to zero at low temperatures. 

To determine the extent of the SANS for wave vector transfer perpendicular to the plane spanned by $(111)$ and $(1,-1,0)$, we performed rocking scans (varying the cryostat rotation angle about the vertical axis), tracking the intensity of the different rods as a function of $Q_\perp$, wave vector transfer perpendicular to the high symmetry plane.  
Figure \ref{flo:SANS_FC-ZFC}(d) shows the $(111)$ scattering has a HWHM of $\Delta Q_{\perp}= 1.6(2) \times 10^{-3} $\AA$^{-1}$ corresponding to a correlation length $\xi_\perp=1/Q_{\perp} = 620(60)$ \AA.  
This evidences  domains walls  extending perpendicular to the  $(111)$ crystallographic axis. Within an angular range of 16$^{\circ}$ no cryostat rotation angular dependence of the intensity was found along $(2,0,\pm 1)$.  The corresponding $Q_{\perp}$ plot is in Fig. \ref{flo:SANS_FC-ZFC}(d) and provides an upper bound of 310(30) \AA\ on correlations perpendicular to the plane spanned by (111) and (1,-1,0). 
Since $(1 \bar10)$ is parallel to the vertical rotation axis the corresponding rocking curve was not measurable in this experimental setup.  Thus we cannot determine whether this scattering is plane-like or rod-like in 3d $Q$-space.

\begin{figure}
	\centering\includegraphics[scale=0.48]{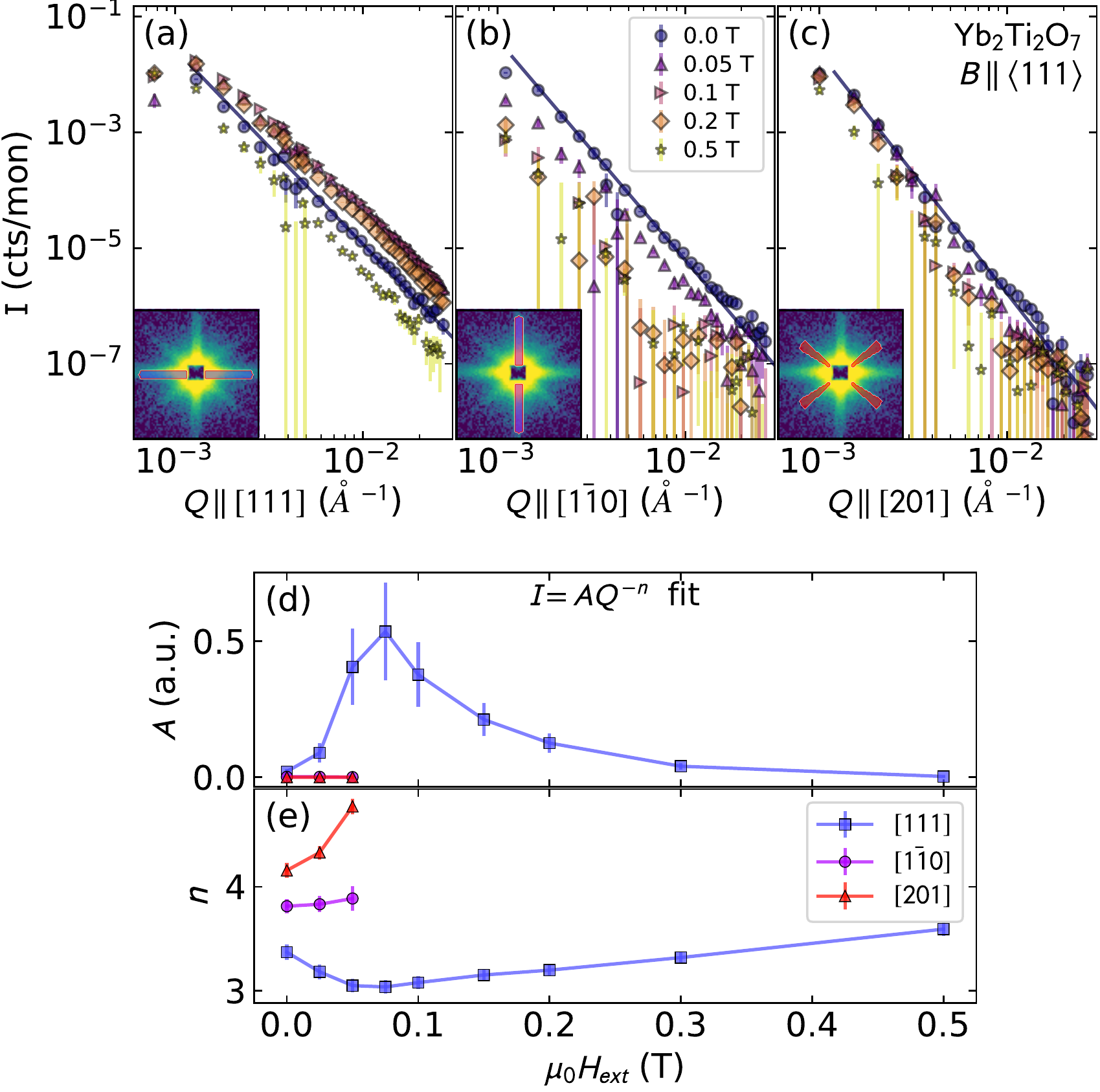}
	
	\caption{Fits to SANS in $\rm Yb_2Ti_2O_7$. (a)-(c) show the $Q$ dependent scattering at various fields in the (f) $(111)$, (g) $(1 \bar 1 0)$, and (h) $(201)$ directions. The regions defining these directions are shown in the insets to frames (a)-(c). (d) and (e) show the results of a Porod exponent fit $I=AQ^{-n}$.
	}
	\label{flo:SANS_Porod}
\end{figure}

Figure \ref{flo:SANS_B-T} shows the temperature and field dependence of the different scattering features [the windows defining the regions of interest are shown in the insets of Fig. \ref{flo:SANS_Porod}(a)-(c)]. The temperature dependence of the zero-field SANS shows a clear onset at the ordering temperature of 270 mK. At 0.1 T and 0.2 T, the ordering transition, shown by the onset of $(111)$ scattering, increases in temperature in accord with the reentrant phase diagram \cite{Scheie2017}. In applied fields of 0.1 T and 0.2 T $(1 \bar10)$ and $(201)$ scattering is absent within the ordered phase, but appears briefly near $T_c$. The paramagnetic, critical, and ordered scattering patterns are shown in Fig. \ref{flo:SI_critical}. 

The field-dependent scattering pattern is shown in Fig. \ref{flo:SANS_B-T}(d)-(f), and the field-dependence from a ZFC state is depicted in Fig. \ref{flo:SANS_B-T}(g)-(j).
At 120 mK (within the ordered phase), an applied field initially enhances the $( 111 )$ scattering and then gradually suppresses it [shown most clearly in Fig. \ref{flo:SANS_B-T}(g)-(j)], while the $( 1 \bar10 )$ and $( 201 )$ scattering is completely suppressed by 0.1 T (where internal demagnetizing field becomes nonzero \cite{Scheie2017}). Reducing the field from 0.8 T at 120 mK results in a much more gradual increase in SANS, with an anomaly in the field dependence of the  $(111)$ intensity when intensity at $(1 \bar10)$ and $(201 )$ reappears at 0.1 T. At 300 mK (where there is no zero field order and the ordered phase is reentrant versus field \cite{Scheie2017}), the intensity at  $(1 \bar10)$ and $(201)$ appears only at the lower field boundary of the ordered phase while hysteresis in the field dependence of the intensity at $(111)$ is much less pronounced than at 100 mK.


The $Q$-dependence of SANS is indicative of magnetic structure on the 100-1000 \angstrom ~length scale \cite{teixeira1988small,Martin1987}. 
The streaks of scattering can be fit to a power law $I=A Q^{-n}$ (where $I$ is scattering intensity, $Q$ is the scattering vector, and $A$ is a fitted constant), which is different for the three different visible streaks as shown in Fig. \ref{flo:SANS_Porod}. These data were taken only from the maximum intensity cryostat rotation angle. In the $(111)$ direction, the exponent is  $n=3.37(7)$ in zero-field and $n=3.04(6)$ at 0.075 T. In the $(1 \bar 10)$ and $(201)$ directions where the SANS vanishes beyond 0.05 T, $n=3.81(7)$ and $4.16(7)$ respectively in zero-field.

For randomly oriented surfaces, the power $n$ is called a Porod exponent and it provides insights into the real space structure under underlying the SANS \cite{Martin1987}. However, this theory does not directly apply to the anisotropic SANS pattern observed in this case. Here we focus on the SANS extending along the (111) direction, which we have shown is rod like in reciprocal space and therefore is associated with planar structures extending perpendicular to (111). 
According to ref. \cite{Sinha1988}, the SANS scattering associated with a sharp discontinuity in the scattering length density normal to $z$ (in this case a ferromagnetic domain wall) with the incident beam along $x$ goes as 
\begin{equation}
S({\bf Q}) = C \frac{1}{Q_z^2} \delta(Q_x) \delta(Q_y).
\end{equation}
Integrating over detector pixels in the direction perpendicular to the plane yields 
\begin{equation}
I = I_0 \int \sin (2\theta) d(2 \theta) d\phi ~ S({\bf Q}) \propto \frac{1}{Q_z^3} 
\end{equation}
\cite{Sinha2019}.
This explains the behavior of the $(111)$ scattering rod: well-separated domain walls in a lamellar pattern with their normals along $ \langle 111 \rangle$.

The streaks of scattering extending  along $(1 \bar1 0)$ and $(201)$ have exponents closer to $n=4$.
$Q^{-4}$ and $Q^{-3}$ SANS power law exponents along different crystallographic directions have been observed from superalloy grain boundaries \cite{bellet1992small}. Similar effects could be at play here: the surfaces of the domain walls where multiple domains meet could easily have either sharp edges or double-curvature, both of which will lead to $Q^{-4}$ scattering \cite{bellet1992small}. This may also explain why the $(201)$ streak intensity is independent of $Q_{\perp}$.  [Fig. \ref{flo:SANS_FC-ZFC}(d)]: if they are due to domain wall edges (not lamellar flat domain walls), they will not have a rod shape and will not have a dramatic angular dependence like $(111)$.

\begin{figure}
	\centering\includegraphics[scale=0.22]{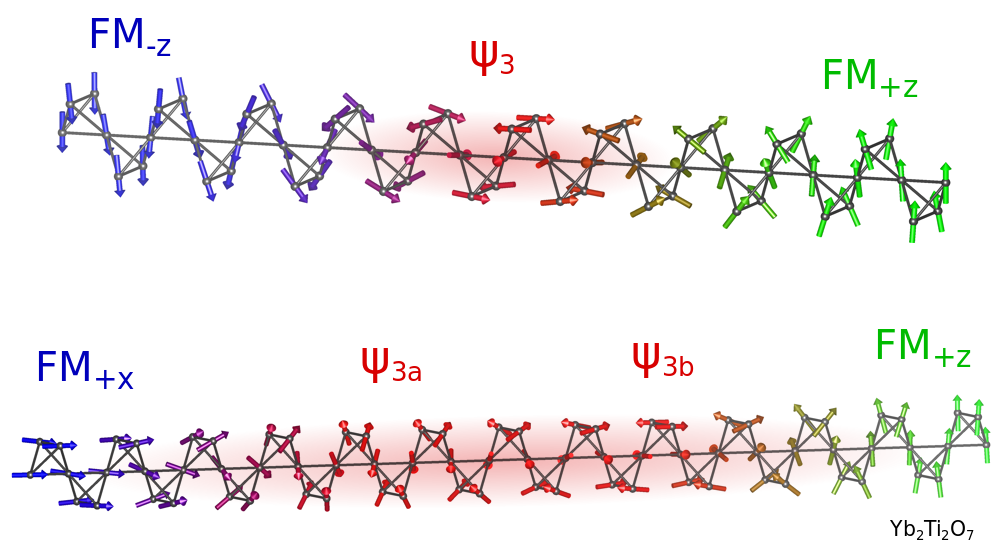}
	
	\caption{Domain walls in $\rm Yb_2Ti_2O_7$ rotating through a $\psi_3$ antiferromagnetic phase. The top image shows a domain wall between $-z$ and $+z$ ferromagnetic domains, and the bottom image shows a domain wall between $+x$ and $+z$, rotating between two states in the $\psi_3$ manifold.
	}
	\label{flo:DomainWall}
\end{figure}

Given these measurements it is important to ask \textit{why} the magnetic domain wall scattering is so strongly anisotropic. 
The extreme anisotropy of the SANS features is unusual for ferromagnets \cite{Michels_2014}.
The answer may lie in the nature of the domain walls. 
Because the magnetic Hamiltonian for $\rm Yb_2Ti_2O_7$ is on the boundary between a canted ferromagnetic and $\psi_3$ antiferromagnetic phase, 
then the lowest energy domain wall includes a slab of $\psi_3$ antiferromagnetic phase, as depicted in Fig. \ref{flo:DomainWall}. For a domain wall separating $+\hat z$ from a $- \hat z$  oriented magnetization domain, the rotation is a simple transition through a $\psi_3$ phase. But if the rotation is to a domain at $90^{\circ}$, the domain wall is a transition to a $\psi_3$ phase, then a rotation within the $\psi_3 - \psi_2$ manifold (which is $U_1$ degenerate at the mean-field level \cite{Savary2012}), and then a transition to the ferromagnetic phase again. On the FM+AFM phase boundary, such a rotation costs zero energy, but slightly away from the phase boundary it is still the lowest-energy way to connect two FM states (see Appendix \ref{app:DomainWalls}).

This observation may explain the dramatic anisotropy in the domain wall structure. The FM domains in $\rm Yb_2Ti_2O_7$ may point in $\pm(100)$, $\pm(010)$, or $\pm(001)$ directions. Dipolar interactions favor domain walls with zero net magnetic charge so that  the magnetization should form the same dot product with the normal to the domain wall on both sides of it. Thus if two domains are magnetized along (100) and (010) then the domain wall may be in the (110) or (111) plane, which may explain the streaks of scattering extending in those directions. Note however, the different power law describing the Q-dependence of the scattering along these two types of streaks.  The (201) feature, however, would support charged domain walls and have a different orgin (evidenced by the lack of angular dependence).

At the FM/AFM boundary, FM and $\psi_3$ become degenerate; and at the mean field level (neglecting thermal and quantum fluctuations), $\psi_3$ and $\psi_2$ are degenerate. This allows for both $\psi_3$ and $\psi_2$ to be present and favored  as domain walls by dipolar energies within a ferromagnetic ordered state.
Thus, AFM regions may be stabilized by dipolar interactions.

To examine whether the apparent  phase coexistence indicated by the inelastic scattering could arise  from domain  walls, we carried out the MC+LSWT analysis for domain walls: we minimized the energy for a particular AFM domain wall and calculated the spin wave spectrum. However, the levels of AFM scattering were far too weak for this to be a reasonable explanation for the broad low-field spectrum; the domain walls were simply too sparse. The observed scattering pattern requires more extended AFM within the sample, which would perhaps be produced by the influence of magnetic dipolar interactions stabilizing broader domain walls or regions of AFM. To analyze this will require inclusion of dipolar interactions in the MC simulation.

\section{Discussion}

This study resolves several long-standing puzzles concerning $\rm Yb_2Ti_2O_7$. First, we have shown that the ground state order is the 2-in-2-out canted ferromagnetism as predicted by theory \cite{Yan2017}. Second, we have shown that $\rm Yb_2Ti_2O_7$ does have dispersive excitations, albeit damped, in the zero-field state. Third, we have shown that these excitations are nearly gapless with an interesting flat-band excitation associated with the $(220)$ peak. Fourth, we have shown that soft zero-field spin wave modes are those of the AFM $\Gamma_5$ phase, which constitutes evidence that $\rm Yb_2Ti_2O_7$ is near the phase boundary and contains a significant volume fraction of short range AFM order within the otherwise FM ground state.

Our SANS data shows a highly anisotropic magnetic domain structure, which may be associated with incorporation of AFM slabs in domain walls.
We also have demonstrated using Monte Carlo simulation that $\psi_2$ is a metastable phase of  $\rm Yb_2Ti_2O_7$ at low temperatures, and fluctuations may be generating the AFM scattering patterns that we observe in low energy inelastic magnetic neutron scattering. 

The nature of the boundary between the FM and AFM phases, as revealed by our spin wave calculations, is quite interesting. 
The linear spin wave calculation shows that the FM ground states have a tendency to fluctuate into the $\Gamma_5$ states. The soft mode at $(220)$ matches the $\Gamma_5$ order parameter and the spectrum gap scales as the square root of the energy cost of the $\Gamma_5$ states. This continuous gap closing suggests a continuous phase boundary between FM and AFM states. However, numerical calculations in ref. \cite{Robert2015} show it is a first order boundary even at finite temperature. These calculations are both approximate, and we leave this apparent contradiction to be resolved in a future study.



The natural question that arises from our calculations and procedure is: why does averaging over the structure factor of different spin configurations reproduce the neutron spectrum (Fig. \ref{flo:SW_FM-AFM-MC})? 
The MC simulations suggest that regions of AFM order are kinetically trapped in a system which has partly lost ergodicity. However, the diffraction results in Fig. \ref{flo:Refinement} show <10\% of the elastic magnetic scattering is in the form of  antiferromagnetic Bragg scattering. This means that the AFM components are localized in space and/or time. The MC simulations do not reveal small AFM regions (smaller than $8\times 8 \times 8$ unit cells). However, these simulations do not include magnetic dipolar interactions which could produce a dense network of AFM regions in between FM domains as a means of reducing the magnetostatic energy.
Under this hypothesis, the flat modes above $(220)$ and $(111)$ may be due to standing spin waves in a finite sized AFM region favored by dipole interactions. If the antiferromagnetism is restricted to domain walls (extended in two dimensions and constrained in a third), its spin waves orthogonal to the wall will have nodes at the edges of the AFM domain, which leads to standing wave resonance modes at nonzero energies. In other words,  the flat modes above $(220)$ and $(111)$ may be spin wave resonance modes in an effective "quantum well".

Alternatively, it could be that $\rm Yb_2Ti_2O_7$ fluctuates in time in and out of the FM and AFM phases, such that AFM scattering appears only at finite energy transfer.
Such fluctuations would occur via quantum effects, which are also neglected by the MC simulations. We know from refs. \cite{Scheie2017,Changlani2017quantum} that quantum effects are important - they can change $T_c$ by a factor of 2 in zero field. 
Moreover if one looks at the classical energy histogram, the energy difference between the AFM and FM manifold 
is small, it is about 0.0025 meV per site. This means that quantum mechanical tunneling locally between the FM and AFM will be possible. 
The classical order parameters describing the system, FM, $\psi_2$, and $\psi_3$ are associated with noncommuting operators at the quantum level, indicating tunneling between these classically defined ordered states will occur. 
If quantum tunneling from one state to another happens coherently in a region on some long time scale ($\sim 0.1$~ns), by measuring the classical order parameter, we might get a sense of domain wall sweeping through the region.
In addition, the flat modes above $(220)$ and $(111)$ may be low-energy modes of the spins tunneling in and out of the AFM phase. Such a mechanism would yield a neutron scattering signal, and these flat modes are one of the most glaring features not captured by semiclassical theory. It must also be added that even the static correlations at 50 mK (i.e. in the ordered phase) show signatures of both FM and AFM correlations, as has been discussed in recent work by Pandey et al~\cite{pandey2019analytical}. These static correlations were modelled by taking a FM-AFM ratio of 2/3 and averaging their individual structure factors similar in spirit to what has been done in our work. Remarkably,  the averaging procedure appears to even account for the \emph{dynamical structure} factor.   
This might be phenomenologically justified through a separation of time scales; the tunnelling being slow enough to accommodate the much faster spin wave excitations.

Many new questions are raised by this study. First, it is not clear why the SANS pattern has such extreme anisotropy, and why domain walls prefer to align normal to the $\langle 111 \rangle$ direction. Large-box classical simulations of AFM domain walls based on the Hamiltonian derived above show a mild preference for a domain to be along $\langle 100 \rangle$, but not $\langle 111 \rangle$ as indicated by the data---but these simulations neglected dipolar interactions which are crucial in domain wall stabilization. 
Second, the broadened zero-field spectrum requires a more complete theoretical explanation. The MC+LSWT spectrum reproduces many features, but the match is not perfect, indicating processes not captured by linear spin wave theory.
Third, the mechanism for phase coexistence needs to be determined. Based on diffraction we can rule out extended pockets of AFM, which leaves (i) domain wall AFM stabilized by dipolar interactions and (ii) dynamic fluctuations into the AFM phase. The specific mechanism needs to be clarified with quantitative theory and further experimental work for example at very high energy resolution.

\section{Conclusion}
	

This study puts the enigmatic pyrochlore $\rm Yb_2Ti_2O_7$ in a new light: one of phase coexistence. Interactions in $\rm Yb_2Ti_2O_7$ create a mixed magnetic state that includes regions of AFM within the otherwise FM ground state, constrained to be finite in space and/or time. Previous studies have speculated about phase coexistence in the paramagnetic phase \cite{pandey2019analytical}. Mutiple lines of evidence in our work show that coexistence indeed occurs  in the ordered phase.  
Many of $\rm Yb_2Ti_2O_7$'s puzzling properties may arise from  coexistence of ferromagnetism with antiferromagnetism. A compound so finely tuned to the phase boundary seems very improbable; perhaps a principle or mechanism remains to be discovered that places it there. But it could also be that in the course of seeking unusual magnetism in myriad frustrated magnets we have finally come across a compound with the unlikely set of interaction parameters that lead to a near degeneracy between FM and AFM. Either way $\rm Yb_2Ti_2O_7$ realizes a unique low $T$ state of matter where ferromagnetism and antiferromagnetism coexist in harmony. 

\section*{Acknowledgments}
This work was supported as part of the Institute for Quantum Matter, an Energy Frontier Research Center funded by the U.S. Department of Energy, Office of Science, Basic Energy Sciences under Award No. DE-SC0019331. This research used resources at the Spallation Neutron Source, a DOE Office of Science User Facility operated by the Oak Ridge National Laboratory. AS and CB were supported through the Gordon and Betty Moore foundation under the EPIQS program GBMF-4532. HJC  was  supported  by  start-up  funds  from  Florida  State  University  and the National High Magnetic Field Laboratory.  HJC thanks the Research Computing Cluster (RCC) at Florida State University  and  XSEDE  (allocation DMR190020) for computational resources. The  National  High  Magnetic  Field  Laboratory is supported by the National Science Foundation through NSF/DMR-1644779 and the state of Florida. We acknowledge helpful discussions with Lisa Debeer-Schmidt, Ken Littrell, Roderich Moessner, Jeff Rau, Nic Shannon, Sunil Sinha, and Oleg Tchernyshyov.

%
%

\quad


\appendix
 
\renewcommand{\thefigure}{A\arabic{figure}}
\renewcommand{\thetable}{A\arabic{table}}
\setcounter{figure}{0}

\section*{Appendices}

\section{Elastic scattering intensities}\label{app:ElasticScattering}

\begin{figure}
	\centering\includegraphics[scale=0.43]{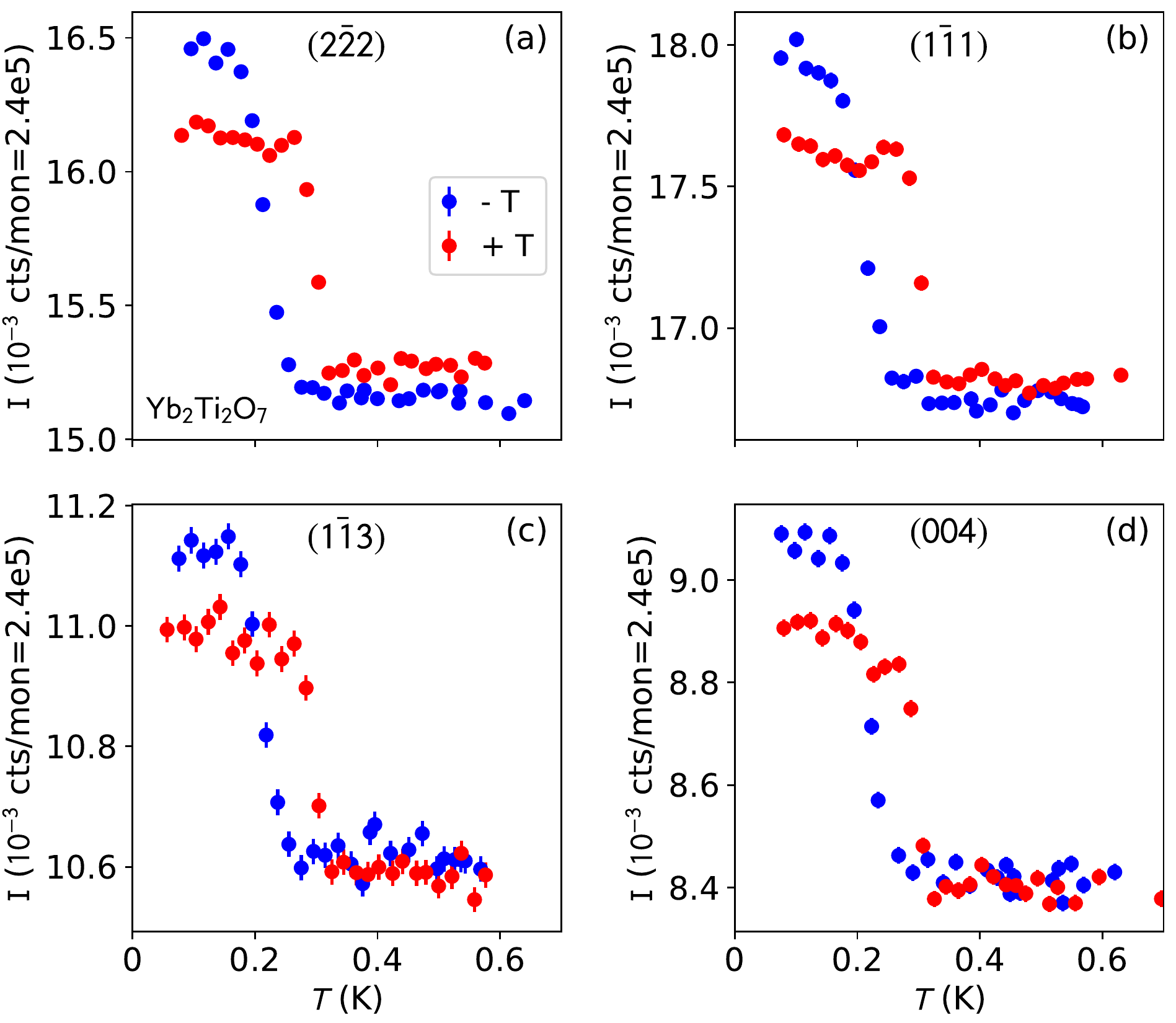}
	
	\caption{Temperature dependence of the $\rm Yb_2Ti_2O_7$ for the $(222)$, $(111)$, $(113)$, and $(004)$ Bragg peaks, measured on the TASP spectrometer with 5.0 meV neutrons. Cooling and heating rates were 3 mK/min, and between the cooling and the heating, a 1 T field was applied. The intensities refined in the main text were the zero-field intensities.}
	\label{SI:DiffractionPeaks}
\end{figure}

The elastic scattering intensities used in the magnetic single crystal refinement were taken from temperature scans at $E_i = E_f = 5$ meV, shown in Fig \ref{SI:DiffractionPeaks}. The $(002)$ and $(220)$ peaks were very weak, so these intensities were taken from an A3 rocking scan for $(002)$, and the ref. \cite{Scheie2017} intensity for $(220)$ (the former experiment had much better statistics). 

\subsection{Multiple scattering on the $(002)$ peak}

\begin{figure*}
	\centering\includegraphics[scale=0.48]{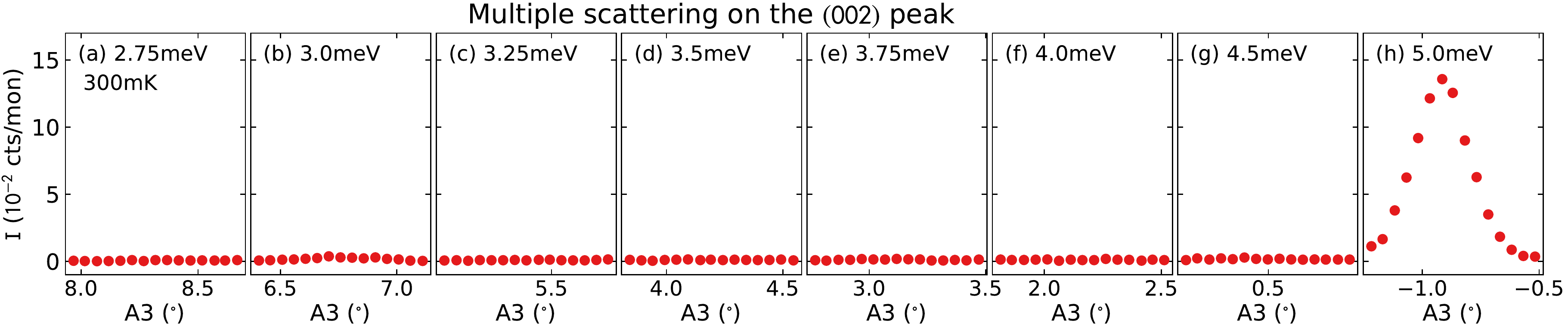}[H]
	
	\caption{Elastic neutron scattering at the $(002)$ peak location in $\rm Yb_2Ti_2O_7$ at 300 mK (above the ordering transition) at various incident energies  with the crystal mounted for diffraction in the $(hhl)$ plane. The only energy at which a peak appears is at 5 meV, which is a clear sign of multiple scattering. Accordingly, all elastic scattering on (002) was performed at $E_i = E_f =4.5$ meV.
	}
	\label{flo:SI_002}
\end{figure*}

As noted in the text, there is discrepancy in previous studies as to whether an $(002)$ magnetic Bragg peak exists in $\rm Yb_2Ti_2O_7$. Part of this discrepancy may be because the $(002)$ intensity is very weak, but multiple scattering  may also play a role.

For the $(002)$ peak mounted with (1,-1,0) perpendicular to the scattering plane and $E_i=E_f=5$ meV neutrons, two multiple scattering pathways exist: $(1 \bar 1 \bar 1) \rightarrow (\bar 1 1 3)$  and $(1 1 1) \rightarrow (\bar 1 \bar 1 1)$. Both of these contribute to the  $(002)$ intensity, as shown in Fig. \ref{flo:SI_002}. Fortunately, this can be remedied by shifting the incident and final neutron energies to 4.5 meV.
However, one of the diffraction studies claiming to see the 2-in-2-out canted ferromagnet ground state used 5 meV neutrons \cite{GaudetRoss_order}, so the refined moment and canting angle from this study are not reliable.

\section{Spin wave fits}\label{app:SpinWaveFits}

\begin{figure}
	\centering\includegraphics[scale=0.2]{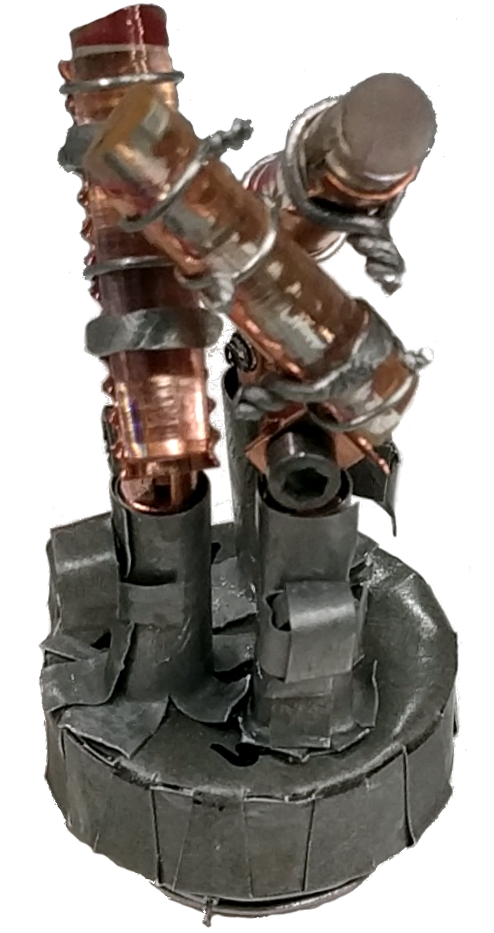}
	
	\caption{Three coaligned $\rm Yb_2Ti_2O_7$ crystals used for the CNCS inelastic neutron scattering experiment. The base of the copper sample holder is wrapped in cadmium.}
	\label{SI:crystals}
\end{figure}

The coaligned crystals used for the CNCS neutron experiment are shown in Fig. \ref{SI:crystals}.

The fits to the 1.5 T data were performed using the $g$ tensor from ref. \cite{Thompson_2017}:
\begin{equation}
g_{\parallel} = 2.14 \>, \quad \quad g_{\perp} = 4.17 .
\end{equation}
We did not fit the $g$ tensor, though doing so  could in principle change the final fitted exchange  parameters. The main point of our analysis was to show first the proximity to the FM+AFM phase boundary, and second to demonstrate the non-uniqueness of a high-field spin wave fit, neither of which depend on the $g$ tensor.

\begin{figure}
	\centering\includegraphics[scale=0.43]{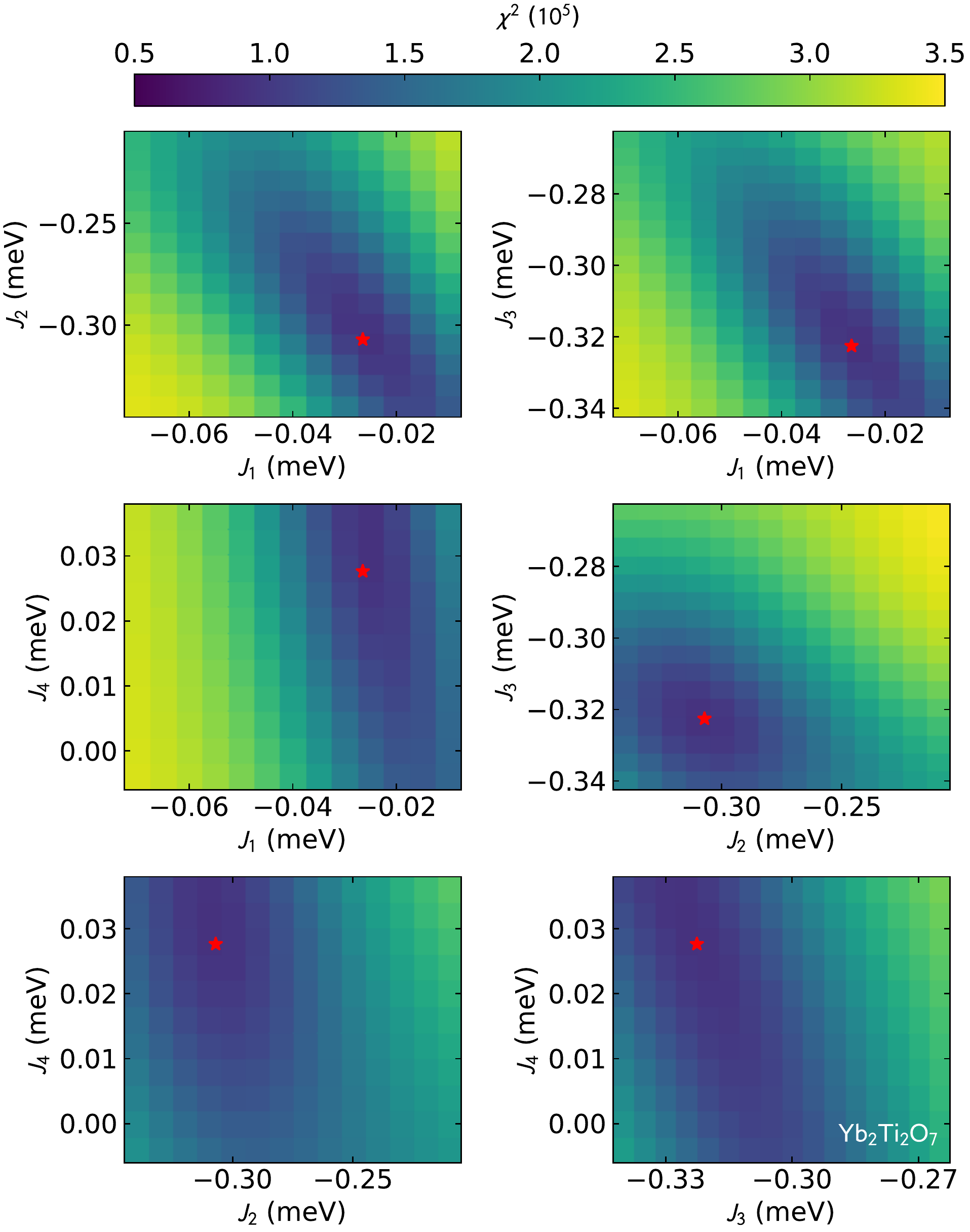}
	
	\caption{Spin wave fit $\chi^2$ to 1.5 T data in $\rm Yb_2Ti_2O_7$ vs. $J_1$, $J_2$, $J_3$, and $J_4$, where the non-plotted dimensions are the best fit value (i.e., when plotting $J_1$ vs $J_2$, $J_3 = -0.322$ meV and $J_4 = 0.028$ meV). The best fit values in eq. \ref{eq:FittedH} are indicated with a red star.}
	\label{SI:Chi2}
\end{figure}

The $\chi^2$ vs. various parameters is plotted in Fig. \ref{SI:Chi2}.
This plot is deceptive because most slices shows a clear minimum $\chi^2$ in parameter space, when in fact there is a line (or rather "cigar") of minimum $\chi^2$ which stretches through this four-dimensional space. Along this line, the 1.5 T scattering is reproduced with similar accuracy, and it was only the zero-field gap which constrained the fit to a point along this line.

\begin{figure}
	\centering\includegraphics[scale=0.38]{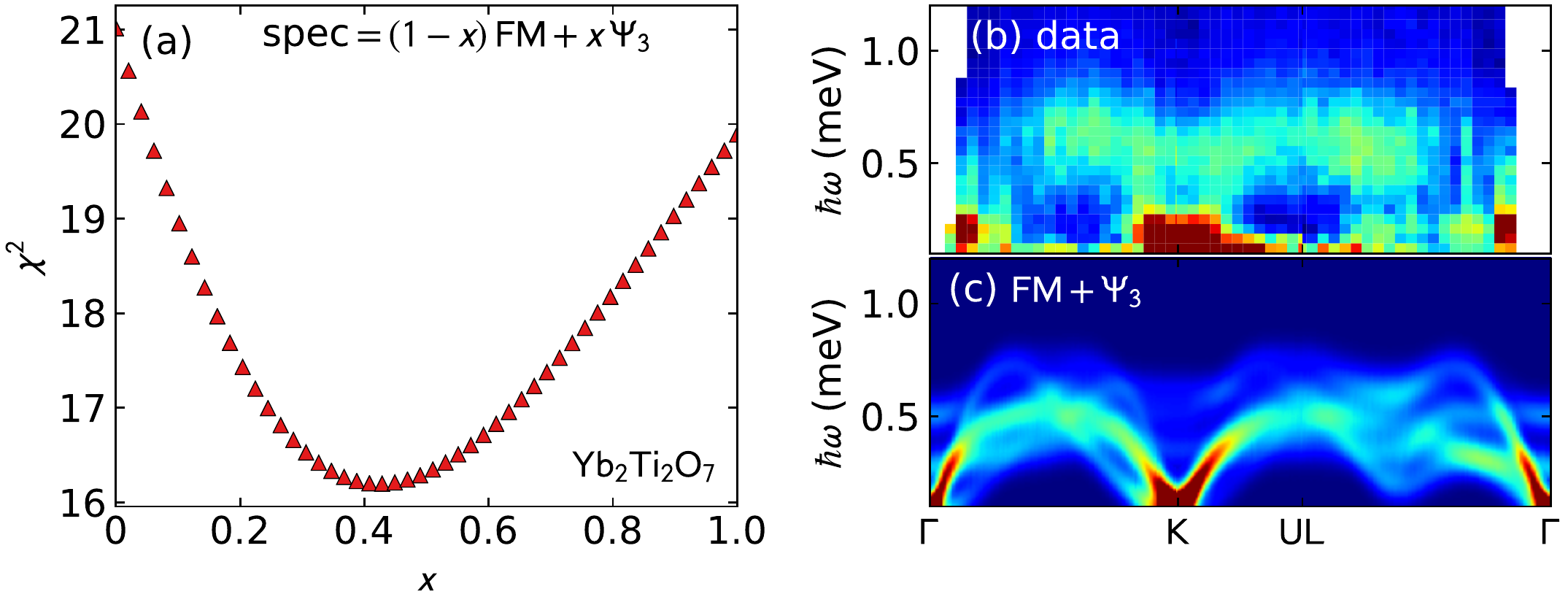}
	
	\caption{Fitted ratio between FM and AFM scattering spectrum. (a) $\chi^2$ of simulation vs. fraction AFM spectrum. (b) Zero field neutron spectrum. (c) Best fit neutron spectrum with x=0.43.}
	\label{SI:FM-AFM-fit}
\end{figure}

The fitted ratio between FM and AFM scattering considering the high-symmetry cuts is shown in Fig. \ref{SI:FM-AFM-fit}. As noted in the text, the best fit value is 43(3)\% AFM.

\section{Soft modes and Bragg peaks of the $\psi_3$ phase}\label{app:SoftModes}

\begin{figure}
	\centering\includegraphics[scale=0.43]{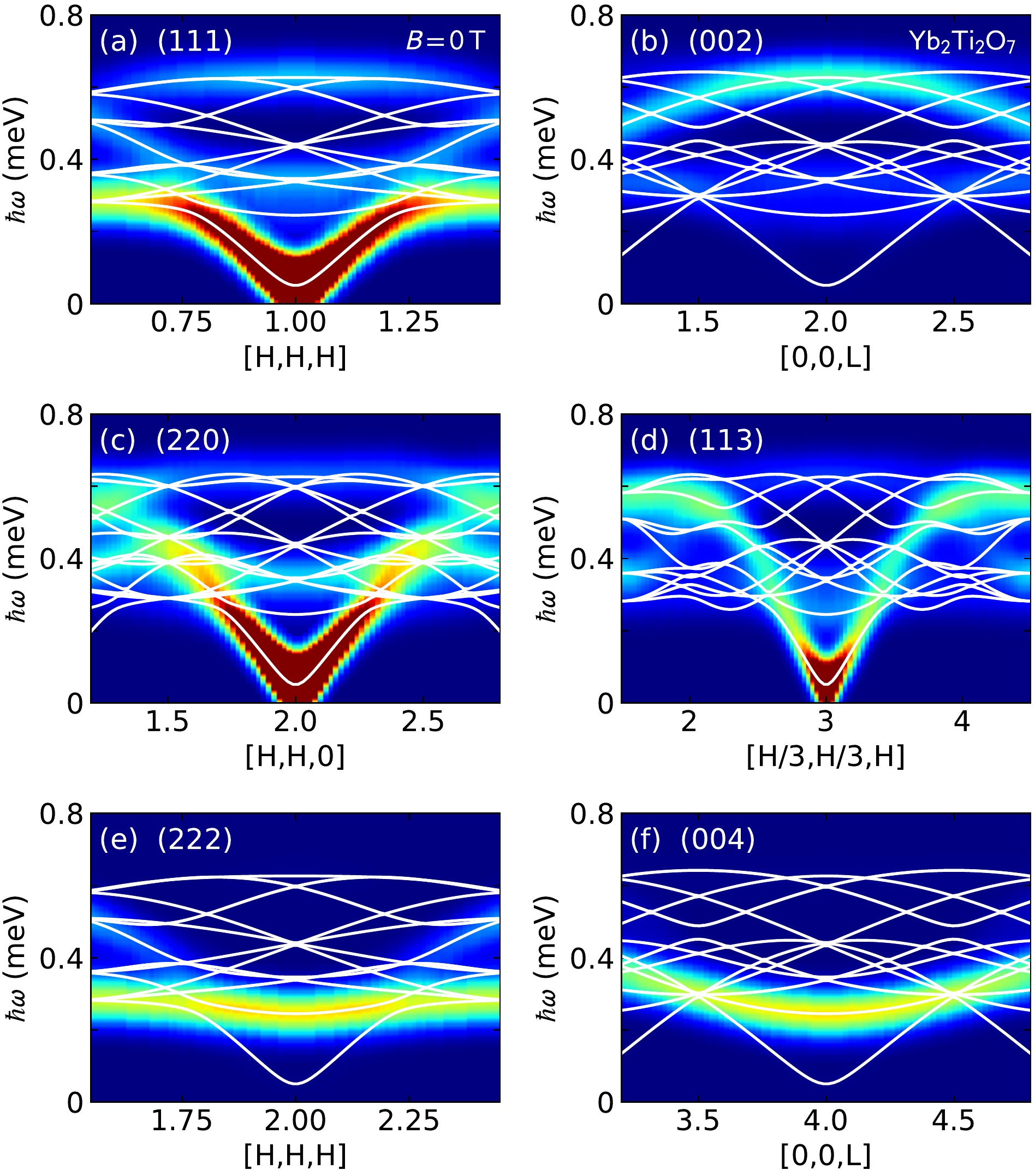}
	
	\caption{Calculated inelastic neutron spectrum around six Bragg peaks in the vicinity of the FM+AFM phase boundary. White lines show the spin wave modes, and the colormap shows the calculated neutron intensity. The only Bragg peaks which have soft modes with nonzero intensity are $(111)$, $(220)$, and $(113)$, which are the magnetic Bragg peaks of the $\psi_3$ antiferromagnetic ordered phase.}
	\label{SI:GapSearch}
\end{figure}

According to linear spin wave theory, as the $\rm Yb_2Ti_2O_7$ Hamiltonian approaches the phase boundary between FM and $\psi_3$ (by tuning $J_2$ for example), soft modes develop above the $(111)$, $(220)$, and $(113)$ Bragg peaks. These soft modes are depicted in Fig. \ref{SI:GapSearch}. As shown in Table \ref{tab:SoftModes}, these Bragg peaks correspond to the Bragg peaks of the $\Gamma_5$ phase which includes both $\psi_3$ and $\psi_2$.

\begin{table}
	
	\caption{Predicted Bragg peaks and soft modes in $\rm Yb_2Ti_2O_7$ for the four magnetic phases in ref. \cite{Yan2017}. The check marks indicate that the Bragg peak is present for that phase.}
	
	\begin{ruledtabular}
		\begin{tabular}{c|c|c|c|c|c}
			Magnetic Bragg Peak & $\psi_2$ & $\psi_3$ & $\psi_4$ & $FM$ & Calculated Soft Modes \tabularnewline
			\hline 
			$(111)$ \hspace{1mm}  1.08 \AA$^{-1}$  & \ding{52} & \ding{52} & \ding{52} & \ding{52} & \ding{52}  \tabularnewline
			$(002)$ \hspace{1mm} 1.25 \AA$^{-1}$  & &  & \ding{52} & \ding{52} &   \tabularnewline
			$(220)$ \hspace{1mm} 1.77 \AA$^{-1}$  & \ding{52} & \ding{52} & \ding{52} & \ding{52} & \ding{52}  \tabularnewline
			$(113)$ \hspace{1mm} 2.07 \AA$^{-1}$  & \ding{52} & \ding{52} & \ding{52} & \ding{52} & \ding{52}  \tabularnewline
			$(222)$ \hspace{1mm} 2.17 \AA$^{-1}$  & &  &  & \ding{52} &   \tabularnewline
			$(004)$ \hspace{1mm} 2.50 \AA$^{-1}$  & &  &  & \ding{52} &   \tabularnewline
		\end{tabular}\end{ruledtabular}
	\label{tab:SoftModes}
\end{table}

\begin{figure*}
	\centering\includegraphics[scale=0.44]{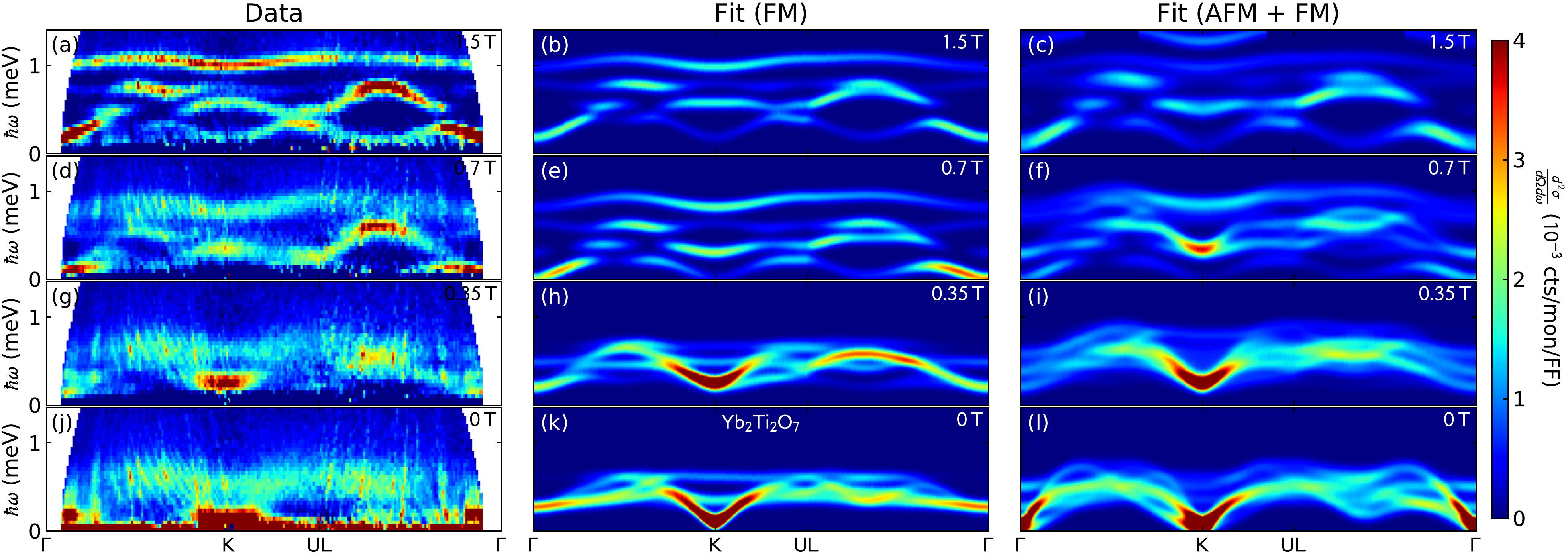}
	
	\caption{Inelastic neutron spectrum of $\rm Yb_2Ti_2O_7$ compared to spin waves from the best fit Hamiltonian and the spin waves from the FM + AFM state. As panels (j)-(l) show, only the zero-field state (and possibly the 0.35 T state) resembles the FM+AFM spectrum.
	}
	\label{flo:SI_FM-AFM}
\end{figure*}

\section{Critical SANS scattering}\label{app:CriticalSANS}

\begin{figure*}
	\centering\includegraphics[scale=0.6]{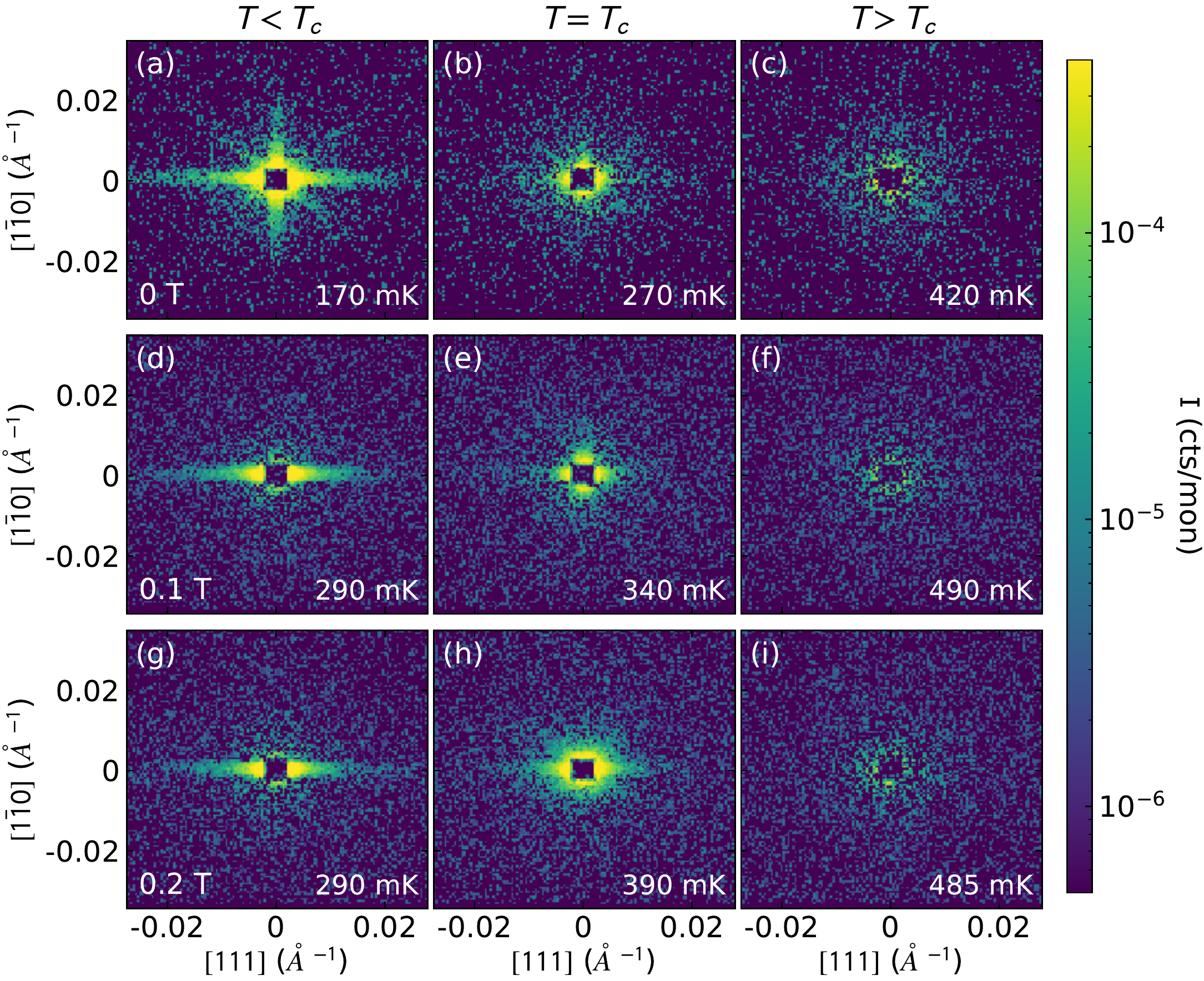}
	
	\caption{SANS intensity for $\rm Yb_2Ti_2O_7$ below, at, and above the critical temperature for three magnetic fields: 0.0 T (a)-(c), 0.1 T (d)-(f), and 0.2 T (g)-(i). Note that the 0.1 T critical scattering in panel (e) shows only vertical and horizontal rods, while the 0.2 T critical scattering in panel (h) shows diagonal $(201)$ rods as well. Data was collected upon cooling.
	}
	\label{flo:SI_critical}
\end{figure*}

As shown in Fig. \ref{flo:SANS_B-T}, there is a brief appearance of $(1 \bar 1 0)$ and $(201)$ scattering at the critical temperature at 0.2 T, but only the $(1 \bar 1 0)$ critical scattering appears at 0.1 T. The paramagnetic, critical, and ordered scattering patterns are shown in Fig. \ref{flo:SI_critical}.

The scattering at the critical temperature in-field, as seen in Fig. \ref{flo:SI_critical}(e) and (h), is different from the scattering in the ordered phase. The in-field ordered scattering shows only a horizontal rod of $(111)$ intensity, while the in-field critical scattering includes $(1 \bar 10)$ intensity, and $(201)$ intensity at 0.2 T. 
This suggests that the domains take on a different structure in the critical regime, and that the structure is different between 0.2 T and 0.1 T, evidenced by the absence of diagonal $(201)$ intensity at 0.1 T.

An alternative to actual critical scattering is SANS associated with domain wall formation. If the domains form in a random state and then reorient along the field direction in a finite amount of time (the $T \sim T_c$ SANS pattern was present for about ten minutes, and the sweep rate was 3.2 mK/min), then this would also produce the temperature dependent scattering we observe. This is also consistent with the incipient rods in Fig. \ref{flo:SI_critical}(e) and (h).

\section{Antiferromagnetic domain walls} \label{app:DomainWalls}

In this appendix we provide a more detailed discussion of antiferromagnetic domain walls. 

We can think of domain walls as a rotational interpolation between different FM ground states. Due to the splayed nature of the $Z_6$ ferromagnetic states, a rotation with respect to a global axis is unable to take one ground state to another. It is easy to find a set of local rotation axes to do so, but a random choice of these local axes will rotate the spins into a high-energy configuration. The proximity to the FM/$\Gamma_5$ boundary provides a natural path to reduce the energy cost of domain walls, as shown in Fig. \ref{flo:DomainWall}. A FM ground state (ex. $+z$) can rotate into the opposite ground state ($-z$) by rotating in to and then out of the corresponding $\psi_3$ state. Note that the magnetic moments of the $\psi_3$ state are orthogonal to those in a ferromagnetic state, which is not true for order parameters of other irreducible representations. Thus this rotation is equivalent to a linear superposition of the FM and $\psi_3$ states, which only cost the same amount of energy as the $\psi_3$ component. 
If the domain wall rotates the magnetization by $90^{\circ}$, the domain wall is a transition to a $\psi_3$ phase, then a rotation within the $\Gamma_5$ AFM manifold (which contains the doublet made of $\psi_2$ and $\psi_3$ states with a continuous $U(1)$ degeneracy \cite{Savary2012}), and then a transition to the ferromagnetic phase again.
In fact, if the exchange parameters are exactly at the phase boundary between SFM and $\psi_3$, the ground state manifold expands to include any linear superposition of the FM and the corresponding $\Gamma_5$ states, as noted by Yan et al \cite{Yan2017}. The domain wall in that case simply explores the ground state manifold and there is no exchange energy costs except for that associated with gradually changing the order parameter. Slightly away from the phase boundary on the FM side the domain walls are favored by dipolar interactions and still incorporate the $\psi_3$ states.

\section{Semiclassical Monte Carlo Simulations}\label{app:SemiclassicalMC}

This section describes the details of the Monte Carlo and linear SWT simulations. 

\subsection{Low energy effective Hamiltonian}
Following refs. \cite{Curnoe_2007,Onoda_2011,Ross_Hamiltonian}, we write the low-energy effective Hamiltonian on the pyrochlore lattice 
with nearest neighbor interactions and Zeeman coupling to an external field ($h=(h_x,h_y,h_z)$) as
\begin{equation}
H = \frac{1}{2} \sum_{ij} J^{\mu\nu}_{ij} S^{\mu}_{i} S^{\nu}_{j} - \mu_{B} h^{\mu} \sum_{i} g^{\mu \nu}_{i} S^{\nu}_{i}
\label{eq:Ham_supp}
\end{equation}
where $i,j$ are nearest neighbors and $\mu,\nu$ refer to $x,y,z$, $S^{\mu}_i$ refer to the spin components at site $i$, and 
$\bf{J}_{ij}$ and $\bf{g}_i$ are bond and site dependent interactions and coupling matrices respectively 
(whose components have been written out in Eq.~\ref{eq:Ham_supp}). The pyrochlore lattice 
has four sublattices which we label as $0,1,2,3$ and we take the relative locations of the sites on a single tetrahedron to be, 
(in units of lattice constant $a$) ${\bf{r}_0}=(1/8,1/8,1/8)$, ${\bf{r}_1}=(1/8,-1/8,-1/8)$, 
${\bf{r}_2}=(-1/8,1/8,-1/8)$ and ${\bf{r}_3}=(-1/8,-1/8,1/8)$. Symmetry considerations dictate that $\bf{J}_{ij}$ and $\bf{g}_{i}$ are 
completely described by four and two scalars respectively. 
$\bf{J}_{ij}$ depends only on the sublattices that $i,j$ belong to
(similarly $\bf{g}_i$ depends only on the sublattice of site $i$), and thus we use the notation 
in terms of $i,j=0,1,2,3$. Also, since $\bf{J}_{ij}=\bf{J}_{ji}^T$, only the $i<j$ matrices are 
written out. The $\bf{J}_{ij}$ matrices are,
\begin{equation}
\bf{J}_{01} \equiv
\left(\begin{array}{ccc}
J_2 & J_4 & J_4 \\
-J_4 & J_1 & J_3 \\
-J_4 & J_3 & J_1 \end{array} \right) 
\bf{J}_{02} \equiv
\left(\begin{array}{ccc}
J_1 & -J_4 & J_3 \\
J_4 & J_2 & J_4 \\
J_3 & -J_4 & J_1 \end{array} \right) \nonumber
\end{equation}
\begin{equation}
\bf{J}_{03} \equiv
\left(\begin{array}{ccc}
J_1 & J_3 & -J_4 \\
J_3 & J_1 & -J_4 \\
J_4 & J_4 & J_2 \end{array} \right) 
\bf{J}_{12} \equiv
\left(\begin{array}{ccc}
J_1 & -J_3 & J_4 \\
-J_3 & J_1 & -J_4 \\
-J_4 & J_4 & J_2 \end{array} \right) \nonumber 
\end{equation}
\begin{equation}
\bf{J}_{13} \equiv
\left(\begin{array}{ccc}
J_1 & J_4 & -J_3 \\
-J_4 & J_2 & J_4 \\
-J_3 & -J_4 & J_1 \end{array} \right) 
\bf{J}_{23} \equiv
\left(\begin{array}{ccc}
J_2 & -J_4 & J_4 \\
J_4 & J_1 & -J_3 \\
-J_4 & -J_3 & J_1 \end{array} \right) 
\end{equation}
Defining $g_{+}=\frac{1}{3}(2g_{xy}+g_{z})$ and $g_{-}=\frac{1}{3}(g_{xy}-g_z)$, the 
$\bf{g}_i$ matrices read as,
\begin{equation}
\bf{g}_{0} \equiv
\left(\begin{array}{ccc}
g_+ & -g_- & -g_- \\
-g_- & g_+ & -g_- \\
-g_- & -g_- & g_+ \end{array} \right) 
\bf{g}_{1} \equiv
\left(\begin{array}{ccc}
g_+ & g_- & g_- \\
g_- & g_+ & -g_- \\
g_- & -g_- & g_+ \end{array} \right) \nonumber
\end{equation}
\begin{equation}
\bf{g}_{2} \equiv
\left(\begin{array}{ccc}
g_+ & g_- & -g_- \\
g_- & g_+ & g_- \\
-g_- & g_- & g_+ \end{array} \right) 
\bf{g}_{3} \equiv
\left(\begin{array}{ccc}
g_+ & -g_- & g_- \\
-g_- & g_+ & g_- \\
g_- & g_- & g_+ \end{array} \right) 
\end{equation}

The interaction part when written in terms of spin directions along the local [111] axes~(denoted by $\Scal$), is,
\begin{eqnarray}
H_{int} &=& \sum_{\langle i,j \rangle}   (2-\lambda) J_{zz} \;\; \Scal^{z}_{i}\Scal^{z}_{j}- \lambda J_{\pm} \Big( \Scal^{+}_{i} \Scal^{-}_{j} + \Scal^{-}_{i} \Scal^{+}_{j} \Big) \nonumber \\ 
& &		  		+ \lambda J_{\pm \pm}\;\;\Big( \gamma_{ij} \Scal^{+}_{i} \Scal^{+}_{j} + \gamma^{*}_{ij} \Scal^{-}_{i} \Scal^{-}_{j} \Big) \nonumber \\ 					 
& &		+ \lambda J_{z,\pm} \;\; \Big[ \Scal^{z}_{i} \Big( \Scal^{+}_{j} \zeta_{ij} + \Scal^{-}_{j} \zeta^{*}_{ij} \Big) + i \leftrightarrow j \Big] \label{eq:Ham}
\end{eqnarray}
where $J_{zz},J_{\pm},J_{\pm\pm},J_{z,\pm}$ are couplings and the parameter $\lambda$ has been introduced by us 
to tune from the classical ice manifold ($\lambda=0$) to real material relevant parameters ($\lambda=1$). 
$\zeta_{ij}$ and $\gamma_{ij}$ are bond dependent phases, 
\begin{equation}
\zeta \equiv
\left(\begin{array}{cccc}
0 & -1 & e^{+i\pi/3} & e^{-i\pi/3} \\
-1 & 0  & e^{-i\pi/3} & e^{+i\pi/3} \\
e^{+i\pi/3} & e^{-i\pi/3} & 0 & -1  \\
e^{-i\pi/3} & e^{+i\pi/3} & -1 & 0 \end{array} \right) 
\gamma = \zeta^{*}
\end{equation}

The relations between $J_1,J_2,J_3,J_4$ and  $J_{zz},J_{\pm},J_{\pm\pm},J_{z,\pm}$ are
\begin{eqnarray}
J_{zz}     &=& -\frac{1}{3}         (+2 J_{1} - J_{2} + 2 J_{3} +  4 J_{4})     \nonumber \\ 
J_{\pm}    &=& +\frac{1}{6}         (+2 J_{1} - J_{2} -   J_{3} -  2 J_{4})     \nonumber \\ 
J_{\pm\pm} &=& +\frac{1}{6}         (+  J_{1} + J_{2} - 2 J_{3} +  2 J_{4})     \nonumber \\ 
J_{z\pm}   &=& +\frac{1}{3\sqrt{2}} (+  J_{1} + J_{2} +   J_{3} -    J_{4})     
\end{eqnarray}
Table \ref{tab:parameters} summarizes the parameters that were used for the calculations  and that were obtained in previous studies in both notations. 

\begin{table*}[htpb]
	\begin{center}
		\begin{tabular}{|c|c|c|c|c||c|c|c|c||c|c|}
			\hline
			Parameter set     &  $J_1$(meV)  & $J_2$ (meV) & $J_3$ (meV) & $J_4$ (meV) &  $J_{zz}$ (meV) & $J_{\pm}$ (meV) & $J_{z\pm}$ (meV) & $J_{\pm \pm}$ (meV) & $\;\;g_{xy}\;\;$ & $\;\;g_{z}\;\;$  \tabularnewline
			\hline
			Ross ~\cite{Ross_Hamiltonian}  &  -0.09 & -0.22 & -0.29 & +0.01 &  0.17   & 0.05   & -0.14 & 0.05  &  4.32   & 1.8    \tabularnewline
			Thompson ~\cite{Thompson_2017}&  -0.028 & -0.326 & -0.272 & +0.049 &  0.026  & 0.074  & -0.159 & 0.048 &  4.17   & 2.14   \tabularnewline
			Robert ~\cite{Robert2015}&  -0.03 & -0.32 & -0.28 & +0.02 &  0.07  & 0.085  & -0.15 & 0.04 &  ?   & ?   \tabularnewline
			This study &  -0.026(2) & -0.307(3) & -0.322(3) & +0.028(2) & 0.094   & 0.087  & -0.161 &  0.0611 & 4.17   & 2.14   \tabularnewline
			\hline
		\end{tabular}
		\caption{Spin Hamiltonian parameter sets used in the paper in two notations for $\rm Yb_2Ti_2O_7$. The reported parameters 
			from one notation were directly converted to the other notation (without accounting for error bars) 
			unless already provided in the reference.}
		\label{tab:parameters}
	\end{center}
\end{table*}

\subsection{Monte Carlo Ground States}\label{app:MonteCarloGroundStates}

\begin{figure}
	\centering\includegraphics[width=0.47\textwidth]{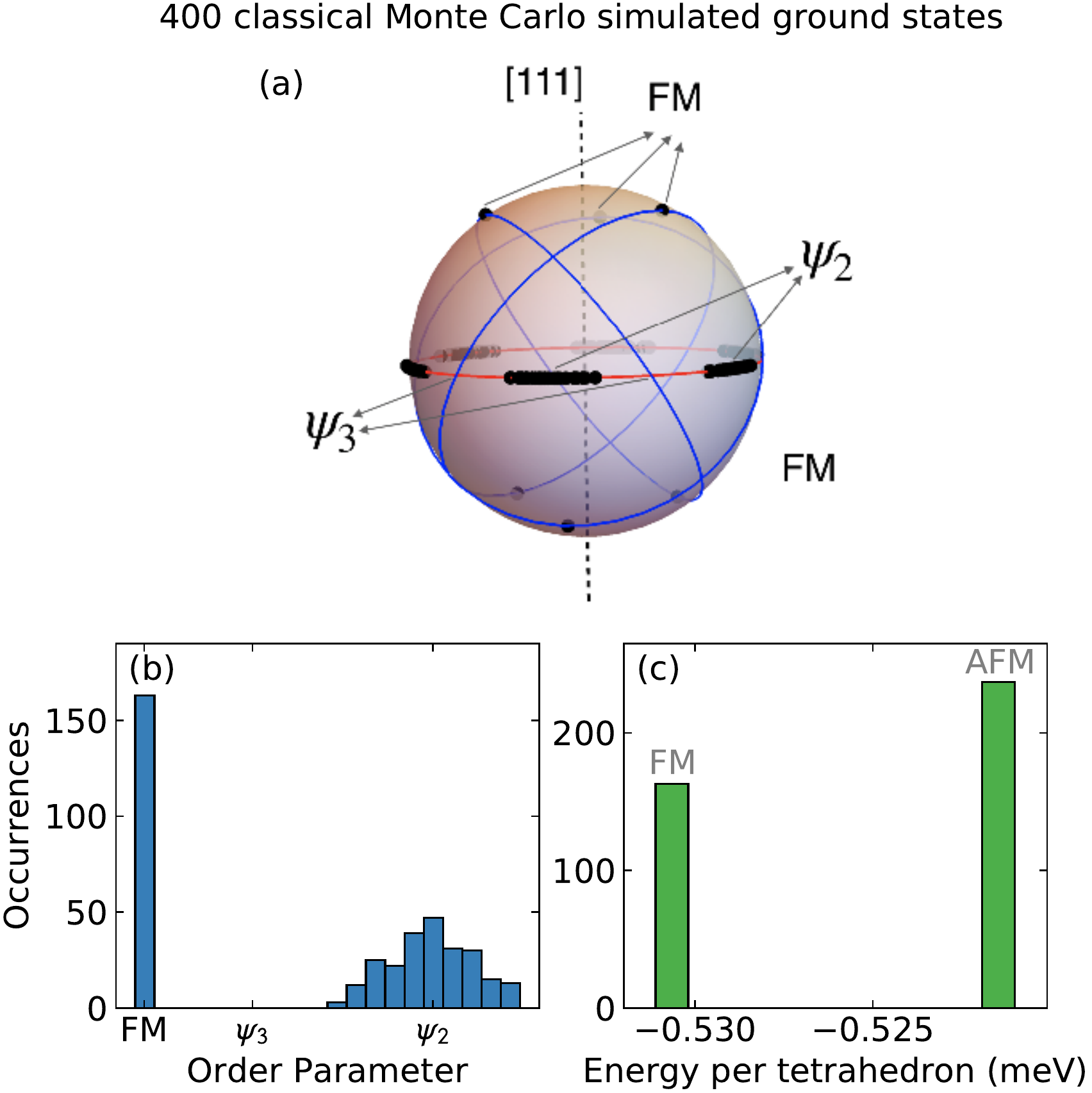}
	
	\caption{Resulting ground states from classical Monte-Carlo simulations. (a) The distribution of ground states (represented by black circles) in a 3D order parameter space (following ref. \cite{Yan2017}). The red line shows the $\Gamma_5$ manifold containing $\psi_2$ and $\psi_3$, and the blue lines show the connections to the FM states which become degenerate at the phase boundary. (b) Histogram of ground states, where the $x$ axis is the line connecting $\psi_2$ and FM ground states through $\psi_3$ (in panel (a) starting at FM, following the blue line until it intersects with the red line and then following it to $\psi_2$). (c) Energy histogram of ground states, showing a bimodal distribution in energy with 40.7\% of the states in a ferromagnetic phase.}
	\label{flo:MC_results}
\end{figure}

In analyzing the Monte Carlo results, we used the order parameters of the different pyrochlore phases \cite{Yan2017}
\begin{widetext}
	\begin{eqnarray}
	m_x &=& \frac{1}{2} (S_0^{x} + S_1^{x} + S_2^{x} + S_3^{x}) \\
	m_y &=& \frac{1}{2} (S_0^{y} + S_1^{y} + S_2^{y} + S_3^{y}) \\ 
	m_z &=& \frac{1}{2} (S_0^{z} + S_1^{z} + S_2^{z} + S_3^{z}) \\
	\psi_2 &=& \frac{1}{2\sqrt{6}} (-2S_0^{x} + S_0^{y} + S_0^{z}  - 2S_1^{x} - S_1^{y} - S_1^{z} + 2S_2^{x} + S_2^{y} - S_2^{z}+2S_3^{x} - S_3^{y} + S_3^{z}) \\
	\psi_3 &=& \frac{1}{2\sqrt{2}} (-S_0^{y} + S_0^{z} + S_1^{y} - S_1^{z} - S_2^{y} - S_2^{z} + S_3^{y} + S_3^{z})
	\end{eqnarray}
\end{widetext}

If the exchange parameters are exactly at the phase boundary between FM and $\Gamma_5$, we can sketch the degenerate ground states as shown in Fig.~\ref{flo:MC_results}(a). The dashed line is the local [111] axis. The red circle is the degenerate $\Gamma_5$ manifold, which (if order by disorder is not considered) is degenerate regardless of the choice of exchange parameters. The blue circles are defined by two FM (for example, with magnetization along $+x$ and $-x$) and their corresponding $\psi_3$ states (that are $\pi$ apart, for example 12 o'clock and 6 o'clock). They correspond to each other in the sense that one can rotate into another with proper choice of local rotation axis without changing the total energy of the spin configuration. In this particular case, there are two zero modes around the $\psi_3$ state: fluctuations along the red or blue circle do not cost energy, while around FM and $\psi_2$ there is only one zero mode. For $\rm Yb_2Ti_2O_7$, slightly away from the FM+AFM phase boundary, the energy associated with FM is slightly lower, but this conceptual energy landscape is still roughly correct. If we look at the fluctuations around FM ordered state, the softest mode is along the blue circle. This is indicated by the eigenvector at the bottom of the energy gap of the FM spin wave at the K point---it is exactly the $\psi_3$ state. However, the energy drops continuously along the line connecting  $\psi_3$ and FM, making $\psi_3$ a saddle point and unstable. $\psi_2$, however, is protected by a finite barrier: it still has a zero mode, but all other local fluctuations cost energy. Thus, although $\psi_2$ and $\psi_3$ are clasically degenerate, $\psi_2$ is a metastable local minimum while $\psi_3$ is an unstable local minimum.  

\subsection{Computing the  spin wave spectrum}

For each of these configurations, we took the single tetrahedron configuration and performed linear spin wave theory to 
get four bands for $\omega(k)$. Correspondingly, $S(q,\omega)$ was also calculated. We follow the notation and formalism in Ross et al. \cite{Ross_Hamiltonian} to obtain the quadratic Hamiltonian (linear in the spin length $s$). We use the sublattice index $a,b$ etc. First, define an orthonormal coordinate system (locally) consisting of ${\bf u}_a, {\bf v}_a$, and ${\bf w}_a$. ${\bf u}_a$ is the direction of the spin, and ${\bf v}_a$ is chosen to be
\begin{equation}
{\bf v}_a = {\bf u}_a \times (1 1 1)/|{\bf u}_a \times (1 1 1)|.
\end{equation} 
Finally
\begin{equation}
{\bf w}_a = {\bf u}_a \times {\bf v}_a.
\end{equation}

One can characterize the deviations along the direction of the spin and perpendicular to it,
\begin{eqnarray}
{\bf S} \cdot {\bf u}_a &=& s - n_a \\
{\bf S} \cdot {\bf v}_a &=& \sqrt{s} x_a \\ 
{\bf S} \cdot {\bf w}_a &=& \sqrt{s} y_a 
\end{eqnarray}
where $n_a = \frac{x_a^2 + y_a^2}{2} - \frac{1}{2}$. We then get the linear spin wave Hamiltonian, 
\begin{equation}
H = (X_{-k}^{T} Y_{-k}^{T})  
\left(\begin{array}{cc}
A_k & C_k \\
C_k^T & B_k \end{array} \right) 
\left(\begin{array}{c}
X_k \\
Y_k \end{array} \right)
\end{equation}
where $(X^T Y^T) = (x_0, ....x_3, y_0,... y_3)$ 
and
\begin{eqnarray}
A_{ab} &=& s {\bf v}_a \cdot {\bf J}_{ab} \cdot {\bf v}_b - s \delta_{ab} ( {\bf u}_a \cdot \sum_{m} {\bf J}_{am} \cdot {\bf u}_m) \\
B_{ab} &=& s {\bf w}_a \cdot {\bf J}_{ab} \cdot {\bf w}_b - s \delta_{ab} ( {\bf u}_a \cdot \sum_{m} {\bf J}_{am} \cdot {\bf u}_m) \\
C_{ab} &=& s {\bf v}_a \cdot {\bf J}_{ab} \cdot {\bf w}_b.  \\ 
\end{eqnarray}
Furthermore,  $D^{ab}_{k}$ where $D=A,B,C$ is, 
\begin{equation}
D^{ab}_{k} = D_{ab} \cos ({\bf k} \cdot ({\bf r}_b - {\bf r}_a))
\end{equation}

$S(q,\omega)$ was calculated using the following expression, for $N=4$, 
working directly in momentum space (this is related to the method in ref.~\cite{Zhang_2019}, but in momentum space instead of real space). 
Also note that if there are exact degeneracies (not discussed above), there are subtleties - 
one would have to normalize the degenerate subspace to satisfy the choice of normalization condition. 
The formulae above and below assume this issue has been taken care of (see ref. \cite{Maestro_2004}). This issue is not relevant if all the eigenfrequencies are non-degenerate 
\begin{widetext}
	\begin{eqnarray}
	\mathcal S_{\text{quantum}}^{\mu \nu} (\mathbf Q, \omega) 
	&=& -\pi S \eta_a^{\mu} \eta_b^{\nu} \sum_{\alpha=1}^{2N} \delta(\omega-\omega_\alpha) \frac{1}{\psi_\alpha^{\dagger} \Gamma \psi_{\alpha}}
	\Big( \Big[ \psi^{\dagger}_\alpha \Big]_{a} \Big[ \psi_{\alpha} \Big]_{b} \Big) \\
	\end{eqnarray}
\end{widetext}
where $\delta$ is the Dirac distribution, $\eta^{\mu}_{a} = \sum_{\kappa = x,y,z} g_{a}^{\mu\kappa}(v_a^{\kappa} w_a^{\kappa})$, $\omega_\alpha$ 
is the $\alpha$-th eigenvalue of $-\Gamma G$, and $\psi_{\alpha}$  is the corresponding right eigenvector. 

\begin{figure}
	\centering\includegraphics[width=0.48\textwidth]{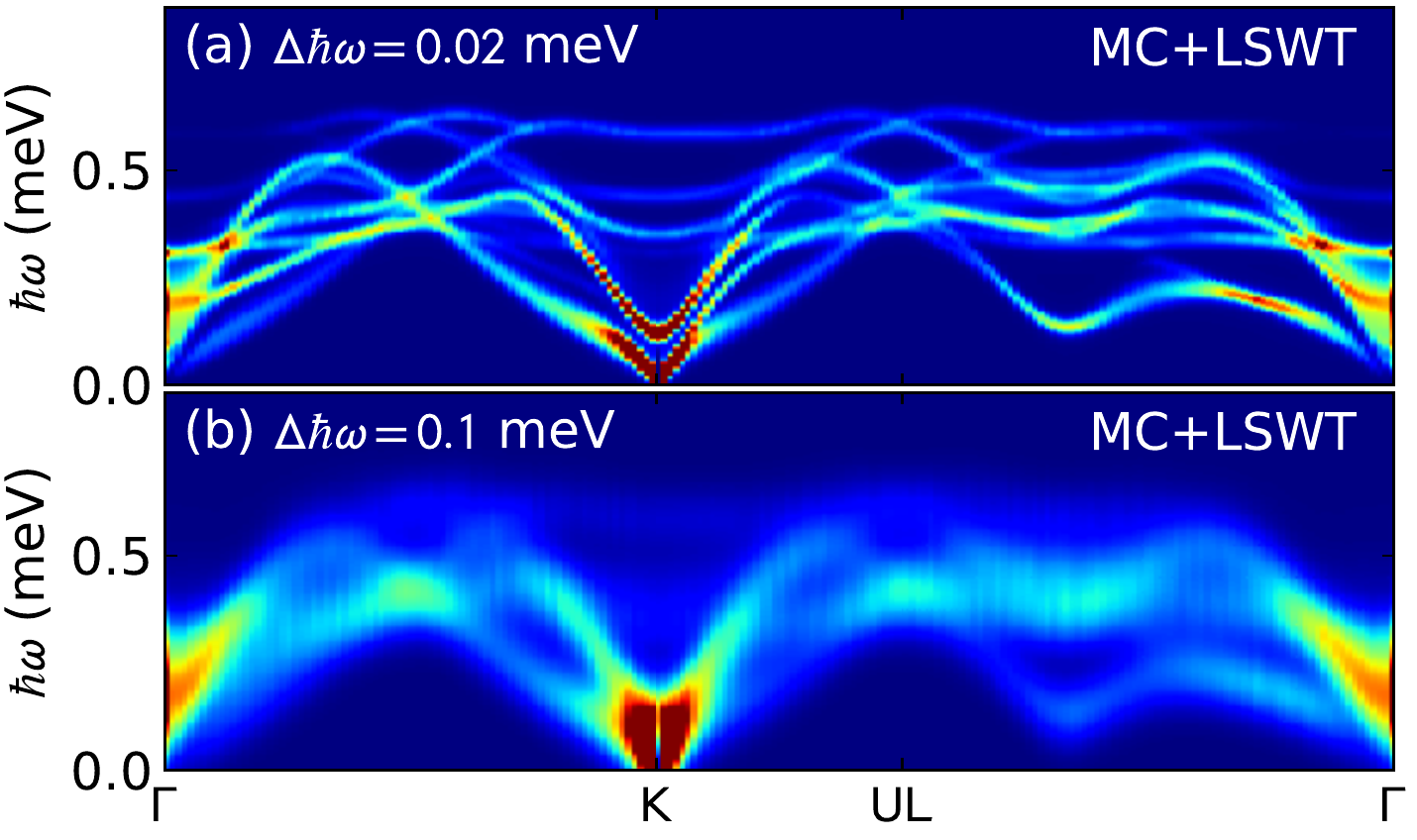}
	
	\caption{High symmetry plots of the predicted scattering from the MC+LSWT simulation. (a) and (b) show the same data, but with different levels of resolution broadening, demonstrating that what appears a continuum in the experiment could be  a multiplicity of spin wave modes.}
	\label{SI:MC_LSWT_broadening}
\end{figure}

In plotting the neutron spectrum, we simulated experimental broadening by convoluting the spectrum in energy with a Gaussian profile with a width defined by the experimental resolution (see Fig. \ref{SI:MC_LSWT_broadening}).


%

\end{document}